\documentclass[
 reprint,
superscriptaddress,
 amsmath,amssymb,
 aps,
]{revtex4-2}

\usepackage{graphicx}
\usepackage{dcolumn}
\usepackage{bm}
\usepackage{siunitx}

\newcommand{\vc}{\mathbf}
\newcommand{\Eq}[1]{Eq.~\eqref{#1}}
\newcommand{\I}{\mathrm{i}}

\newcommand{\beq}{\begin{equation}}
\newcommand{\eeq}{\end{equation}}
\newcommand{\nn}{\nonumber}
\newcommand{\vect}[1]{\mathbf{#1}}

\begin{document}

\preprint{APS/123-QED}

\title{Three-body scattering area of identical bosons in two dimensions}

\author{Junjie Liang}
 
 \email{junjieliang@pku.edu.cn}
 \affiliation{
 International Center for Quantum Materials, School of Physics, Peking University, Beijing 100871, China\\
}

\author{Hongye Yu}
\email{hongye.yu@stonybrook.edu}
\affiliation{
 Department of Physics and Astronomy, State University of New York at Stony Brook, Stony Brook, New York
11794-3800, USA\\
}
 
\author{Shina Tan}
 \email{shinatan@pku.edu.cn}
\affiliation{
 International Center for Quantum Materials, School of Physics, Peking University, Beijing 100871, China\\
}

\date{}

\begin{abstract}
We study the wave function $\phi^{(3)}$ of three identical bosons 
colliding with zero energy, zero total momentum, and zero orbital angular momentum in two dimensions,
interacting with short-range potentials such that the two-body two-dimensional (2D) scattering length $a$ is finite.
We derive the asymptotic expansions of the wave function when the three pairwise distances $s_1$, $s_2$, and $s_3$ are all large (111-expansion)
or when the distance between one pair of particles and the third particle is large (21-expansion), assuming that
 the leading order term  in the 111-expansion grows much slower than $B^n$ at $B\to\infty$
no matter how small the positive constant $n$ is, where
$B=\sqrt{(s_1^2+s_2^2+s_3^2)/2}$ is the hyperradius. At the order $B^{-2}\ln^{-3}(B/a)$ in the $111$-expansion,
we discover a three-body parameter $D$ whose dimension is length squared, and we call it the three-body scattering area.
This scattering area should be contrasted with the three-body scattering area for three bosons with infinite or zero two-body 2D scattering length studied in J. Liang and S. Tan, Phys. Rev. A \textbf{109}, 063328 (2024).
If the two-body interaction is attractive such that it supports two-body bound states,
$D$ acquires a negative imaginary part and we derive the relation between the imaginary part of $D$
and the probability amplitudes for the production of the two-body bound states in the three-body collision.
If the interaction potentials are slightly modified, we derive the shift of $D$ in terms of $\phi^{(3)}$
and the slight changes of the two-body and the three-body potentials.
We also study the effects of $D$ and $\phi^{(3)}$ on the three-body and the
many-body physics, including the three-body ground state energy in a large periodic volume,
the many-body energy and the three-body correlation function of the dilute 2D Bose gas,
and the three-body recombination rates of the 2D ultracold atomic Bose gases.
\end{abstract}

\maketitle

\section{Introduction}
Dimensionality is crucial to the properties of the quantum systems. In two dimensions, quantum and thermal fluctuations are particularly significant. Ultracold atomic gases provide a highly controllable platform for exploring the two-dimensional (2D) quantum physics~\cite{safonov1998observation,PhysRevLett.87.130402,Hadzibabic_2011}. The interaction effects of these gases are governed by few-body low-energy collisions, from which one can extract a few important parameters. Usually, the two-body $s$-wave collisions dominate the interaction effects in ultracold atomic gases. For two particles with equal mass $m$, scattering with a low energy $\hbar^2 k_E^2/m$ (where $k_E>0$) in the center-of-mass frame in two dimensions, the cotangent of the $s$-wave scattering phase shift $\delta_s$ can be expanded as~\cite{kagan1982quasi,BJVerhaar_1984}
\begin{equation}\label{phase shift}
    \cot \delta_s =\frac{2}{\pi}\ln\frac{k_Ea}{2e^{-\gamma}}+\frac{r_s}{2}k_E^2+O(k_E^4),
\end{equation}
where $\gamma=0.5772\ldots$ is Euler's constant, $a$ is the two-body 2D $s$-wave scattering length, and $r_s$ is the two-body 2D $s$-wave effective range (the dimension of $r_s$ is length squared).

For a sufficiently dilute 2D Bose gas such that $\rho a^2$ is a small parameter, where $\rho$ is the number density of the 2D gas, Schick first demonstrated that the ground state energy per particle can be expanded in terms of $1/\ln(\rho a^2)$~\cite{schick1971two}.   Schick derived the leading order term (mean-field term) of the ground state energy of the 2D Bose gas~\cite{schick1971two}.
In some subsequent studies, people derived the beyond-mean-field terms that are universal in the sense that they only depend on $\rho$ and $a$ \cite{popov1972theory,lozovik1978ground,fisher1988dilute,kolomeisky1992renormalization,ovchinnikov1993description,lieb2001rigorous,al2002low,pricoupenko2004variational,pilati2005quantum,mora2009ground,pastukhov2019ground,fournais2024ground}. However, if the Bose gas is not sufficiently dilute, nonuniversal effects may become significant, causing thermodynamic quantities to depend on other few-body low-energy scattering parameters. The nonuniversal effect of the effective range  $r_s$ has been explored in Refs. \cite{PhysRevA.81.013612,beane2010ground,PhysRevLett.118.130402,beane2018effective,Tononi_2018,condmat4010020,PhysRevA.107.033325}. Additionally, the three-body coupling can play a crucial role \cite{beane2010ground,PhysRevLett.110.145301,PhysRevLett.115.075303,PhysRevA.100.042707,PhysRevA.109.063328}. In the resonant regime such that $\left|\ln\left(\rho a^2\right)\right|\sim 1$,  three-body effects may be comparable to two-body effects~\cite{PhysRevLett.110.145301}. Notably, if the three-body effective interaction is repulsive, it may stabilize the 2D dipolar Bose system~\cite{PhysRevLett.115.075303}. When the two-body interaction is tuned such that $a=\infty$ or $0$, the three-body effect dominates~\cite{PhysRevA.100.042707,PhysRevA.109.063328}.

The two-body low-energy scattering parameters, such as the $a$ and $r_s$ in \Eq{phase shift}, can be introduced through the study of the two-body wave function at low energy. Similarly, the three-body low-energy scattering parameters can be defined. One of us \cite{tan2008three} defined the three-body scattering hypervolume for three identical bosons in three dimensions. This concept has been generalized to other systems involving three particles with unequal masses in three dimensions \cite{wang2021three}, 
three spin-polarized fermions in three dimensions \cite{wang2021scattering}, two dimensions \cite{wang2022scattering}, and one dimension \cite{wang2023three}, three identical bosons with either infinite or zero two-body 2D scattering length in two dimensions \cite{PhysRevA.109.063328}, three two-component fermions in three dimensions \cite{PhysRevA.111.053313}, as well as three identical bosons with total orbital angular momentum quantum number $L=2$ in three dimensions \cite{ip2025three}. 
The three-body parameter quantifies the strength of the low-energy effective three-body interaction, so it is a fundamental parameter in the few-body and many-body physics in the ultracold regime (with thermal de Broglie wave lengths far exceeding the range of interaction).

In this paper, we first study three identical bosons with a finite two-body 2D scattering length $a$ in two dimensions.
We derive the asymptotic expansions of the wave function $\phi^{(3)}$ of three bosons colliding with zero energy, zero total momentum, and zero orbital angular momentum, at large interparticle distances, from which we define the three-body scattering area $D$. 
If the two-body interaction is attractive such that it supports two-body bound states,
$D$ acquires a negative imaginary part and we derive the relation between the imaginary part of $D$
and the probability amplitudes for the production of the two-body bound states in the three-body collision.
If the interaction potentials are slightly modified, we derive the shift of $D$ in terms of $\phi^{(3)}$
and the slight changes of the interaction potentials.
We then find approximate formulas for the adiabatic shift of the three-body ground state energy in a large periodic volume due to a slight change of $D$,
the three-body nonlocal correlation function of the dilute 2D Bose gas in terms of $\phi^{(3)}$, 
 the adiabatic shift of the many-body energy of the dilute 2D Bose gas due to a slight change of $D$,
 and the three-body recombination rate constants of the 2D ultracold atomic Bose gases in terms of the imaginary part of $D$ at various temperatures.

\section{Asymptotic expansions of the three-body wave function\label{sec:asymptotics}}
We consider three identical bosons with position vectors $\vect r_1$, $\vect r_2$, and $\vect r_3$ on the 2D plane, interacting with the two-body potential $\frac{\hbar^2}{m}U(\vc{r}_1,\vc{r}_2,\vc{r}_1',\vc{r}_2')$ and the three-body potential $\frac{\hbar^2}{m}U(\vc{r}_1,\vc{r}_2,\vc{r}_3,\vc{r}_1',\vc{r}_2',\vc{r}_3')$. 
We define a Cartesian coordinate system on the plane,
such that the position vector $\vect r_i$ of the $i$th boson is expanded as $\vect r_i=r_{ix}\hat{\vect x}+r_{iy}\hat{\vect y}$, where $\hat{\vect x}$ and $\hat{\vect y}$ are unit vectors long the $+x$ axis and the $+y$ axis respectively, and $r_{ix}$ and $r_{iy}$ are the Cartesian components of $\vect r_i$. 

We define a reflection operator $P_{\hat{\vc{n}}}$ for any 2D unit vector $\hat{\vc{n}}$, such that
\beq\label{Pn}
P_{\hat{\vc{n}}}\vc{r}_i\equiv \vc{r}_i-2\left(\vc{r}_i\cdot{\hat{\vc{n}}}\right){\hat{\vc{n}}}.
\eeq
In particular, 
\beq\label{Py}
P_{\hat{\vc{y}}}\vc{r}_i\equiv r_{ix}\hat{\vect x}-r_{iy}\hat{\vect y}.
\eeq

We assume the following properties for the two-body potential $\frac{\hbar^2}{m}U(\vc{r}_1,\vc{r}_2,\vc{r}_1',\vc{r}_2')$
and the three-body potential $\frac{\hbar^2}{m}U(\vc{r}_1,\vc{r}_2,\vc{r}_3,\vc{r}_1',\vc{r}_2',\vc{r}_3')$ \cite{PhysRevA.109.063328}. 
\begin{itemize}
  \item Spatial translational invariance:  for any 2D vector $\Delta\vect r$,
$U(\vc{r}_1+\Delta\vect r,\vc{r}_2+\Delta\vect r,\vc{r}_1'+\Delta\vect r,\vc{r}_2'+\Delta\vect r)=U(\vc{r}_1,\vc{r}_2,\vc{r}_1',\vc{r}_2')$, and $U(\vc{r}_1+\Delta\vect r,\vc{r}_2+\Delta\vect r,\vc{r}_3+\Delta\vect r,\vc{r}_1'+\Delta\vect r,\vc{r}_2'+\Delta\vect r,\vc{r}_3'+\Delta\vect r)=U(\vc{r}_1,\vc{r}_2,\vc{r}_3,\vc{r}_1',\vc{r}_2',\vc{r}_3')$.
\item Finite range:
the interaction is limited within a finite inter-particle distance $r_e$, such that
$U(\vc{r}_1,\vc{r}_2,\vc{r}_1',\vc{r}_2')=0$ if $|\vc{r}_2-\vc{r}_1|>r_e$ or $|\vc{r}_2'-\vc{r}_1'|>r_e$.
Similarly, $U(\vc{r}_1,\vc{r}_2,\vc{r}_3,\vc{r}_1',\vc{r}_2',\vc{r}_3')=0$ if $|\vc{r}_i-\vc{r}_j|>r_e$ or $|\vc{r}_i'-\vc{r}_j'|>r_e$ for any $i,j\in \{1,2,3\}$.
  \item Bosonic symmetry: because of Bose statistics, we can symmetrize $U$ without losing generality: $U(\vc{r}_1,\vc{r}_2,\vc{r}_1',\vc{r}_2')=U(\vc{r}_2,\vc{r}_1,\vc{r}_1',\vc{r}_2')=U(\vc{r}_1,\vc{r}_2,\vc{r}_2',\vc{r}_1')$.
  Similarly, $U(\vc{r}_1,\vc{r}_2,\vc{r}_3,\vc{r}_1',\vc{r}_2',\vc{r}_3')$ is symmetric under the interchange of $\vc{r}_i$ and $\vect r_j$, and symmetric under the interchange of $\vc{r}_i'$ and $\vect r_j'$, where $i,j\in\{1,2,3\}$.
\item Hermiticity:  $U(\vc{r}_1,\vc{r}_2,\vc{r}_1',\vc{r}_2')=U^*(\vc{r}_2',\vc{r}_1',\vc{r}_2,\vc{r}_1)$,
and $U(\vc{r}_1,\vc{r}_2,\vc{r}_3,\vc{r}_1',\vc{r}_2',\vc{r}_3')=U^*(\vc{r}_3',\vc{r}_2',\vc{r}_1',\vc{r}_3,\vc{r}_2,\vc{r}_1)$.
  \item Rotational invariance: for any rotation $R$ on the 2D plane, $U(\vc{r}_1,\vc{r}_2,\vc{r}_1',\vc{r}_2')=U(R\vc{r}_1,R\vc{r}_2,R\vc{r}_1',R\vc{r}_2')$ and $U(\vc{r}_1,\vc{r}_2,\vc{r}_3,\vc{r}_1',\vc{r}_2',\vc{r}_3')=U(R\vc{r}_1,R\vc{r}_2,R\vc{r}_3,R\vc{r}_1',R\vc{r}_2',R\vc{r}_3')$.
  \item Galilean invariance: 
  $\vc{r}_1+\vc{r}_2\equiv\vc{r}_1'+\vc{r}_2'$ for $U(\vc{r}_1,\vc{r}_2,\vc{r}_1',\vc{r}_2')$, implying that only three of the four vectors $\vc{r}_1$, $\vc{r}_2$, $\vc{r}_1'$, and $\vc{r}_2'$ are independent. $\vc{r}_1+\vc{r}_2+\vc{r}_3\equiv\vc{r}_1'+\vc{r}_2'+\vc{r}_3'$  for $U(\vc{r}_1,\vc{r}_2,\vc{r}_3,\vc{r}_1',\vc{r}_2',\vc{r}_3')$, implying that only five of the six vectors $\vc{r}_1$, $\vc{r}_2$,  $\vc{r}_3$, $\vc{r}_1'$, $\vc{r}_2'$, and $\vc{r}_3'$ are independent.
  \item $\hat{\vc{y}}$-reversal invariance:
  $U(\vc{r}_1,\vc{r}_2,\vc{r}_1',\vc{r}_2')=U(P_{\hat{\vc{y}}}\vc{r}_1,P_{\hat{\vc{y}}}\vc{r}_2,P_{\hat{\vc{y}}}\vc{r}_1',P_{\hat{\vc{y}}}\vc{r}_2')$
  and $U(\vc{r}_1,\vc{r}_2,\vc{r}_3,\vc{r}_1',\vc{r}_2',\vc{r}_3')=U(P_{\hat{\vc{y}}}\vc{r}_1,P_{\hat{\vc{y}}}\vc{r}_2,P_{\hat{\vc{y}}}\vc{r}_3,P_{\hat{\vc{y}}}\vc{r}_1',P_{\hat{\vc{y}}}\vc{r}_2',P_{\hat{\vc{y}}}\vc{r}_3')$.
  Since the interactions are rotationally invariant, the ${\hat{\vc{y}}}$-reversal invariance is equivalent to the $\hat{\vc{n}}$-reversal invariance for any 2D unit vector $\hat{\vc{n}}$.
\end{itemize}

We define the following Jacobi coordinates \cite{tan2008three}:
\begin{subequations}
\begin{equation}
    \vc{s}_1=\vc{r}_2-\vc{r}_3,\quad \vc{s}_2=\vc{r}_3-\vc{r}_1,\quad
    \vc{s}_3=\vc{r}_1-\vc{r}_2,
\end{equation}
\begin{eqnarray}
    \vc{R}_1=\vc{r}_1-\frac{1}{2}(\vc{r}_2+\vc{r}_3),\nonumber\\
    \vc{R}_2=\vc{r}_2-\frac{1}{2}(\vc{r}_3+\vc{r}_1),\nonumber\\
    \vc{R}_3=\vc{r}_3-\frac{1}{2}(\vc{r}_1+\vc{r}_2).
\end{eqnarray}
We also define the hyperradius \cite{tan2008three}
\begin{equation}
    B\equiv\sqrt{(s_1^2+s_2^2+s_3^2)/2}=\sqrt{R_i^2+\frac{3}{4}s_i^2},
\end{equation}
where $i\in\{1,2,3\}$.
\end{subequations}

We consider the three-body scattering wave function $\phi^{(3)}(\vect r_1,\vect r_2,\vect r_3)$ at zero collision energy with the following properties.
\begin{itemize}
    \item Zero total linear momentum and translational invariance: for 
any 2D vector $\Delta \vc{r}$,
    $\phi^{(3)}(\vect r_1+\Delta \vc{r},\vect r_2+\Delta \vc{r},\vect r_3+\Delta \vc{r})=\phi^{(3)}(\vect r_1,\vect r_2,\vect r_3)$.
    \item Zero total orbital angular momentum and rotational invariance: for any rotation $R$ on the 2D plane,  $\phi^{(3)}(R\vect r_1,R\vect r_2 ,R\vect r_3)=\phi^{(3)}(\vect r_1,\vect r_2,\vect r_3)$.
    \item Even $\hat{\vc{y}}$-parity: $\phi^{(3)}(P_{\hat{\vc{y}}}\vect r_1,P_{\hat{\vc{y}}}\vect r_2,P_{\hat{\vc{y}}}\vect r_3)=+\phi^{(3)}(\vect r_1,\vect r_2,\vect r_3)$. Since $\phi^{(3)}(\vect r_1,\vect r_2,\vect r_3)$ is rotationally invariant, the ${\hat{\vc{y}}}$-reversal invariance is equivalent to the $\hat{\vc{n}}$-reversal invariance for any 2D unit vector $\hat{\vc{n}}$: $
\phi^{(3)}(P_{\hat{\vc{n}}}\vect r_1,P_{\hat{\vc{n}}}\vect r_2,P_{\hat{\vc{n}}}\vect r_3)=+\phi^{(3)}(\vect r_1,\vect r_2,\vect r_3)
$.
\end{itemize}
  
$\phi^{(3)}(\vect r_1,\vect r_2,\vect r_3)$ satisfies the Schr\"{o}dinger equation
\begin{widetext}
\begin{align}\label{3-body SE real space}
    &-\frac{1}{2}\left(\nabla_1^2+\nabla_2^2+\nabla_3^2\right)\phi^{(3)}(\vc{r}_1,\vc{r}_2,\vc{r}_3)+\frac{1}{2}\int d^2s_1'U\left(\vc{s}_1,\vc{s}_1'\right)\phi^{(3)}\left(\vc{R}_1,\frac{\vc{s}_1'}{2},-\frac{\vc{s}_1'}{2}\right)\nonumber\\
    &+\frac{1}{2}\int d^2s_2'U\left(\vc{s}_2,\vc{s}_2'\right)\phi^{(3)}\left(-\frac{\vc{s}_2'}{2},\vc{R}_2,\frac{\vc{s}_2'}{2}\right)+\frac{1}{2}\int d^2s_3'U\left(\vc{s}_3,\vc{s}_3'\right)\phi^{(3)}\left(\frac{\vc{s}_3'}{2},-\frac{\vc{s}_3'}{2},\vc{R}_3\right)\nonumber\\
    &+\frac{1}{6}\int d^2r_1'd^2r_2' U(\vc{r}_1,\vc{r}_2,\vc{r}_3,\vc{r}_1',\vc{r}_2',\vc{r}_3')\phi^{(3)}(\vc{r}_1',\vc{r}_2',\vc{r}_3')=0,
\end{align}
\end{widetext}
where $\nabla_i^2$ is the Laplacian with respect to $\vc{r}_i$, and $U(\vc{s},\vc{s}')\equiv U(\vc{s}/2,-\vc{s}/2,\vc{s}'/2,-\vc{s}'/2)$.

Even though we have specified that $\phi^{(3)}$ is the wave function of three bosons with zero energy, zero total momentum, zero orbital angular momentum, and even $\hat{\vect y}$-parity, we have not completed the definition of $\phi^{(3)}$.
There is an \emph{infinite} number of three-body scattering wave functions satisfying these specified properties.
The most important of them, for purposes of studying the ultracold many-body physics, is the one that grows at the \emph{slowest} rate when the three pairwise distances $s_1$, $s_2$, and $s_3$ go to infinity simultaneously.
If there are no interactions, the most important three-boson zero-energy wave function is a constant.
When we introduce interactions gradually, this constant wave function should evolve into a function
that grows at most like a polynomial of $\ln B$ at large pairwise distances (this logarithmic dependence on $B$ is caused by the two-body interactions, since the two-body zero-energy $s$-wave collision wave function behaves like $\ln(s/a)$ when the pairwise distance $s$ is greater than the range $r_e$ of the interaction).
It is this particular three-body wave function that we will study in the following.

Before deriving the asymptotic expansions of $\phi^{(3)}$, we will first define some two-body special functions, which will be important in the 21-expansion (the expansion of $\phi^{(3)}$ in powers of $1/R$, when one pair of bosons are held at a constant distance $s$
but the distance $R$ between the center of mass of the pair and third boson is large).

\subsection{\label{sec:level2}Two-body special functions}
Consider two identical bosons, each of which has mass $m$, interacting with two-body potential $\frac{\hbar^2}{m}U(\vc{s},\vc{s}')$. The two-body Hamiltonian in the center-of-mass frame is $\hbar^2 H_{\vc{s}}/m$, where
\begin{eqnarray}
    H_{\vc{s}}X(\vc{s})\equiv - \nabla^2_\vc{s} X(\vc{s})+\frac{1}{2}\int d^2s'U(\vc{s},\vc{s}')X(\vc{s}').
    \label{HX}
\end{eqnarray}
We define some two-body special functions \cite{tan2008three} characterized by the orbital angular momentum quantum number $l$ and a unit vector $\hat{\vc{n}}$: $\phi^{(l)}_{\hat{\vc{n}}}(\vc{s})$, $f^{(l)}_{\hat{\vc{n}}}(\vc{s})$, $g^{(l)}_{\hat{\vc{n}}}(\vc{s})$, $\ldots$, satisfying
\begin{subequations}\label{Hphi2body}
\begin{align}
H_{\vc{s}}\phi^{(l)}_{\hat{\vc{n}}}(\vc{s})&=0\label{Hphi_l},\\
H_{\vc{s}}f^{(l)}_{\hat{\vc{n}}}(\vc{s})&=\phi^{(l)}_{\hat{\vc{n}}}(\vc{s})\label{Hf_l},\\
H_{\vc{s}}g^{(l)}_{\hat{\vc{n}}}(\vc{s})&=f^{(l)}_{\hat{\vc{n}}}(\vc{s}),\\
\cdots,\nonumber
\end{align}
\end{subequations}
$\phi^{(l)}_{\hat{\vect n}}(\vect s)\propto\cos(l\theta)$ for any fixed $|\vect s|$,
and similarly for $f^{(l)}_{\hat{\vect n}}(\vect s)$ and $g^{(l)}_{\hat{\vect n}}(\vect s)$ etc.,
where $\theta$ is the angle from $\hat{\vect n}$ to $\hat{\vect s}$.

For $s$-wave collisions, $l=0$, the special functions do not depend on $\hat{\vect n}$ and we may simply write
$\phi(\vect s)\equiv\phi^{(0)}_{\hat{\vect n}}(\vect s)$, $f(\vect s)\equiv f^{(0)}_{\hat{\vect n}}(\vect s)$, and $g(\vect s)\equiv g^{(0)}_{\hat{\vect n}}(\vect s)$, etc.
For $d$-wave collisions, $l=2$, we write $\phi^{(d)}_{\hat{\vect n}}(\vect s)\equiv\phi^{(2)}_{\hat{\vect n}}(\vect s)$, etc. 

There is some ``gauge freedom" in the definitions of $\phi^{(l)}_{\hat{\vect n}}(\vect s)$, $f^{(l)}_{\hat{\vc{n}}}(\vc{s})$, and
$g^{(l)}_{\hat{\vc{n}}}(\vc{s})$, etc.: 
$\tilde{\phi}^{(l)}_{\hat{\vect n}}(\vect s)=c_0\phi^{(l)}_{\hat{\vect n}}(\vect s)$,
$\tilde{f}^{(l)}_{\hat{\vc{n}}}(\vc{s})=c_0f^{(l)}_{\hat{\vc{n}}}(\vc{s})+c_1 \phi^{(l)}_{\hat{\vc{n}}}(\vc{s})$, $\tilde{g}^{(l)}_{\hat{\vc{n}}}(\vc{s})=c_0g^{(l)}_{\hat{\vc{n}}}(\vc{s})+c_1 f^{(l)}_{\hat{\vc{n}}}(\vc{s})+c_2 \phi^{(l)}_{\hat{\vc{n}}}(\vc{s})$, $\ldots$, for any coefficients $c_0$, $c_1$, $c_2$, etc., satisfy 
\begin{equation}
H_{\vect s}\tilde{\phi}^{(l)}_{\hat{\vect n}}(\vect s)=0,~~
H_{\vect s} \tilde{f}^{(l)}_{\hat{\vc{n}}}(\vect s)=\tilde{\phi}^{(l)}_{\hat{\vc{n}}}(\vect s),~~H_{\vect s} \tilde{g}^{(l)}_{\hat{\vc{n}}}(\vect s)=\tilde{f}^{(l)}_{\hat{\vc{n}}}(\vect s),
\end{equation}
etc. To fix the gauge freedom, we may specify the explicit formulas for the two-body special functions at $s>r_e$.

If $s>r_e$, the special functions  satisfy 
\begin{subequations}\label{two-body laplace equatioin}
\begin{align}
-\nabla^2_\vc{s}\phi^{(l)}_{\hat{\vc{n}}}(\vc{s})&=0,\\
-\nabla^2_\vc{s}f^{(l)}_{\hat{\vc{n}}}(\vc{s})&=\phi^{(l)}_{\hat{\vc{n}}}(\vc{s}),\\
-\nabla^2_\vc{s}g^{(l)}_{\hat{\vc{n}}}(\vc{s})&=f^{(l)}_{\hat{\vc{n}}}(\vc{s}),\\
\cdots.\nonumber
\end{align}
\end{subequations}

Solving Eqs.~(\ref{two-body laplace equatioin}), we find that if $s>r_e$,
\begin{subequations}\label{phi,s>re}
\begin{align}
    \phi(\vc{s})=&~\ln{\frac{s}{a}}\label{phis,s>re},\\
    f(\vc{s})=&-\frac{\pi r_s}{4}-\frac{s^2\left(\ln{\frac{s}{a}}-1\right)}{4}\label{f,s>re},\\
    \phi^{(d)}_{\hat{\vc{n}}}(\vc{s})=&~\left(\frac{s^2}{8}-\frac{4a_d}{\pi s^2}\right)\cos(2\theta)\label{phidx>re},
\end{align}
\end{subequations}
where we have fixed the gauge freedom in the definitions of $\phi(\vect s)$, $f(\vect s)$, and $\phi^{(d)}_{\hat{\vect n}}(\vect s)$.
The $a_d$ in \Eq{phidx>re} is the $d$-wave scattering ``length" whose dimension is length to the fourth power. 

For a two-body collision with energy $E=\hbar^2k_E^2/m$ in the center-of-mass frame, where $k_E>0$, 
 the $l$-wave wave function can be generally written as
\begin{equation}\label{phi,bessel}
    \phi^{(l)}(\vect s,E)\propto \big[J_l(k_E s)\cot\delta_l-Y_l(k_Es)\big]e^{\pm\mathrm{i}\,l\theta}
\end{equation}
at $s>r_e$, where $\delta_l$ is the $l$-wave scattering phase shift, and $J_l$ and $Y_l$ are the Bessel functions. 
$\cot\delta_s$ satisfies \Eq{phase shift}. More generally \cite{HAMMER2009500}

\begin{equation}
    k_E^{2l}\left[\cot\delta_l-\frac{2}{\pi}\ln\left(k_E\rho_l\right)\right]=-\frac{1}{a_l}+\frac{1}{2}r_l k_E^2+O(k_E^4).
\end{equation}

\subsection{Asymptotics of $\phi^{(3)}(\vc{r}_1,\vc{r}_2,\vc{r}_3)$ at large distances}
Having defined the two-body special functions, we now study the three-body problem.
When the separation $\vect s\equiv\vect s_1$ between particles 2 and 3 is fixed but $\vect R\equiv\vect R_1$ (the displacement from the center of mass of particles 2 and 3 to particle 1) is sufficiently large, \Eq{3-body SE real space} is simplified as
\begin{equation}\label{SE1}
    H_{\vc{s}}\phi^{(3)}(\vc{R},\vc{s}/2,-\vc{s}/2)=\frac{3}{4}\nabla_{\vc{R}}^2\phi^{(3)}(\vc{R},\vc{s}/2,-\vc{s}/2).
\end{equation}
In the limit $R\to\infty$ for any fixed $\vect s$, the three-body wave function can be expanded as
\begin{equation}\label{phi21}
    \phi^{(3)}(\vc{R},\vc{s}/2,-\vc{s}/2)=\sum_{j}^{\infty}S^{(-j)}(\vc{R},\vc{s}),
\end{equation}
where $S^{(-j)}(\vc{R},\vc{s})$ scales roughly like $R^{-j}$ (including the possibility of $R^{-j}\ln^n{R}$ where $n$ is some number). We call \Eq{phi21} the 21-expansion.

When the interpaticle distances $s_1$, $s_2$, and $s_3$ go to infinity simultaneously,  all interaction potentials vanish, and \Eq{3-body SE real space} is reduced to
\begin{equation}\label{SE2}
(\nabla_1^2+\nabla_2^2+\nabla_3^2)\phi^{(3)}(\vc{r}_1,\vc{r}_2,\vc{r}_3)=0.
\end{equation}
In such limit, the three-body wave function can be expanded as
\begin{equation}\label{phi111}
    \phi^{(3)}(\vc{r}_1,\vc{r}_2,\vc{r}_3)=\sum_{i}^{\infty}T^{(-i)}(\vc{r}_1,\vc{r}_2,\vc{r}_3),
\end{equation}
where $T^{(-i)}(\vc{r}_1,\vc{r}_2,\vc{r}_3)$ scales roughly like $B^{-i}$ (including the possibility of $B^{-i}\ln^n{B}$). We call \Eq{phi111} the 111-expansion.

When $R\gg s\gg r_e$, $T^{(-j)}$  and $S^{(-j)}$  can be further expanded as
\begin{subequations}\label{double expansion}
\begin{align}
    T^{(-j)}=\sum_n t^{(n,-j-n)},\label{T double expansion}\\
    S^{(-j)}=\sum_n t^{(-j,n)},
\end{align}
\end{subequations}
where $t^{(i,n)}$ roughly scales like $R^i s^n$ (including the possibility of some extra factors that scale like $\ln^\alpha R\ln^\beta s$).

We assume that $\phi^{(3)}$  scales as $B^0\ln^n B$ when the three bosons are far apart from each other. Therefore, \Eq{phi111} is more explicitly written as
\begin{equation}\label{phi111expansion}
    \phi^{(3)}(\vc{r}_1,\vc{r}_2,\vc{r}_3)=\sum_{i\geq0}T^{(-i)}(\vc{r}_1,\vc{r}_2,\vc{r}_3).
\end{equation}
Therefore, when the distance $s$ between two bosons is fixed, and the distance $R$ between the center of mass of these two bosons and the third boson is large, $\phi^{(3)}$ behaves as $R^0\ln^n R$.
So \Eq{phi21} is more explicitly written as
\begin{equation}\label{phi21expansion}
    \phi^{(3)}(\vc{R},\vc{s}/2,-\vc{s}/2)=\sum_{j\geq 0}S^{(-j)}(\vc{R},\vc{s}).
\end{equation}
Substituting \Eq{phi111expansion} into \Eq{SE2} and expanding both sides of \Eq{SE2} in powers of $1/B$,  we get
\begin{equation}\label{LaplaceT-i}
(\nabla_1^2+\nabla_2^2+\nabla_3^2)T^{(-i)}=0
\end{equation}
if $s_1$, $s_2$, and $s_3$ are all nonzero.
Substituting \Eq{phi21expansion} into \Eq{SE1}, and expanding both sides of \Eq{SE1} in powers of $1/R$, we get
\begin{subequations}
\begin{align}
H_{\vc{s}} S^{(0)}(\vc{R},\vc{s})&=0,\label{HsS0}\\
H_{\vect s}S^{(n)}(\vect R,\vect s)&=0,~~-2<n<0,\label{HsS(-1)}\\
H_{\vect s}S^{(-2)}(\vect R,\vect s)&=\frac34\nabla_{\vect R}^2S^{(0)}(\vect R,\vect s),\label{HsS(-2)}\\
H_{\vect s}S^{(-3)}(\vect R,\vect s)&=\frac34\nabla_{\vect R}^2S^{(-1)}(\vect R,\vect s),
\end{align}
\end{subequations}
and so on.

From Eqs.~\eqref{double expansion}, \eqref{phi111expansion}, and \eqref{phi21expansion}, we get
\begin{equation}
    t^{(i,j)}=0,~~\text{if } i+j>0~\text{or}~i>0. \label{t,i+j>0 or i>0}
\end{equation}

In the following, we derive $T^{(-i)}(\vc{r}_1,\vc{r}_2,\vc{r}_3)$ and $S^{(-j)}(\vc{R},\vc{s})$  step by step.
 
\paragraph*{\textbf{Step 1: Determination of $T^{(0)}$ and $S^{(0)}$.}} 
The solution to the \Eq{HsS0} with even $\hat{\vect y}$-parity is
\begin{equation}\label{S0 1st}
    S^{(0)}(\vc{R},\vc{s})=F^{(0)}(R)\phi(\vc{s})+\sum_{l=2,4,\cdots}X^{(0)}_l(R)\phi^{(l)}_{\hat{\vc{R}}}(\vc{s}),
\end{equation}
where $F^{(0)}(R)$ is a function of the magnitude of $\vc{R}$, and it scales as $R^{0}\ln^{n}{R}$ and $n$ is some number to be determined, and similarly for $X^{(0)}_l(R)$. 
The higher partial wave $\phi^{(l)}_{\hat{\vc{R}}}(\vc{s})$ contains a term proportional to $s^{l}$ at $s>r_e$ [see \Eq{phidx>re}], so if $X^{(0)}_l(R)\neq0$, the expansion of $S^{(0)}(\vc{R},\vc{s})$ at $R\gg s\gg r_e$ would contain a nonzero term $t^{(0,l)}$ with $l\geq 2$, contradicting \Eq{t,i+j>0 or i>0}. Therefore we must have $X^{(0)}_l(R)=0$, and
\begin{equation}\label{S0 2nd}
    S^{(0)}(\vc{R},\vc{s})=F^{(0)}(R)\phi(\vc{s}).
\end{equation}
Expanding the above equation at $R\gg s\gg r_e$, we find
\begin{subequations}
\begin{equation}\label{t(0,0)}
    t^{(0,0)}=F^{(0)}(R)\ln\frac{s}{a},
\end{equation}
\begin{equation}
    t^{(0,j)}=0,\quad j\neq0.
\end{equation}
\end{subequations}

When $R\gg s\gg r_e$, $T^{(0)}$ can be further expanded as
\begin{align}\label{T0 double expansion}
	T^{(0)}
	= t^{(0,0)}+o(R^0).
\end{align}

According to \Eq{LaplaceT-i}, $(\nabla_{\vect s_i}^2+\frac34\nabla_{\vect R_i}^2)T^{(0)}=0$ if $0<s_i\ll R_i$.
Further taking into account \Eq{t(0,0)}, \Eq{T0 double expansion}, and the complete symmetry of $T^{(0)}$ under the exchange of any two bosons, we get
\begin{equation}\label{T0, SE}
    \frac{\nabla_1^2+\nabla_2^2+\nabla_3^2}{2}T^{(0)}=2\pi\sum_{i=1}^3 F^{(0)}(R_i)\delta(\vc{s}_i).
\end{equation}
Using the Green's function to solve the above equation, we get
\begin{equation}\label{T0,first}
    T^{(0)}=\lim_{\Lambda\to\infty}C_{\Lambda}-\frac{1}{2\pi}\sum_{i=1}^3\int_{R'<\Lambda} d^2R'\frac{F^{(0)}(R')}{\frac{3}{4}s_i^2+(\vc{R}_i-\vc{R}')^2},
\end{equation}
where $C_\Lambda$ is a function of $\Lambda$. Expanding \Eq{T0,first} at $R\gg s\gg r_e$ and using \Eq{t(0,0)} and \Eq{T0 double expansion}, we get
\begin{align}\label{F(0) int equ}
    &\lim_{\Lambda\to\infty}\int_0^{\Lambda}F^{(0)}(R')\bigg[\frac{Z_{\sqrt{3}a/4}(R'-R)}{|R'-R|}\frac{1}{R'+R}\nonumber\\
    &+\frac{2}{\sqrt{R^4+R'^4+R^2R'^2}}\bigg]R'dR'-C_\Lambda=0,
\end{align}
where $\frac{Z_{\sqrt{3}a/4}(x)}{|x|}$ is a generalized function called ``Z-function'' \cite{tan2008three}:
\begin{subequations}
\label{Zfunctions1}
\begin{equation}
    \frac{Z_{\sqrt{3}a/4}(x)}{|x|}=\frac{1}{|x|}, \text{ if }x\ne0,
\end{equation}
\begin{equation}
    \int_{|x|<\sqrt{3}a/4}\frac{Z_{\sqrt{3}a/4}(x)}{|x|}dx=0.
\end{equation}
\end{subequations}

Solving \Eq{F(0) int equ}, we get
\begin{align}\label{C(Lambda)}
    C_\Lambda=~&\ln^3\frac{\Lambda}{a}+\frac{9}{2}\left(\ln\frac{4}{3}\right)\ln^2\frac{\Lambda}{a}\nonumber\\
    &+\left(\frac{27}{4}\ln^2\frac{4}{3}-6\omega\right)\ln\frac{\Lambda}{a}-9\omega\ln\frac{4}{3}\nonumber\\
    &+\frac{27}{8}\ln^3\frac{4}{3}-\frac{1}{2}\left(\ln\frac{4}{3}\right)\mathrm{Li}_2\left(\frac{3}{4}\right)\nonumber\\
    &-\mathrm{Li}_3\left(\frac{3}{4}\right)+\frac{3}{2}\zeta(3),
\end{align}
\begin{equation}\label{F0}
    F^{(0)}(R)=\ln^{2}{\widetilde{R}}-2\omega,
\end{equation}
where $\widetilde{R}=\frac{8R}{3\sqrt{3}a}$,
\begin{equation}\label{omega}
\omega=\frac{1}{2}\left(\ln2\right)\left(\ln\frac{4}{3}\right)+\frac{1}{4}\text{Li}_2\left(\frac{3}{4}\right)\approx 0.344320,
\end{equation}
$\zeta(n)=\sum_{k=1}^\infty k^{-n}$ is the Riemann zeta function,
and $\mathrm{Li}_n(z)\equiv\sum_{k=1}^\infty z^k/k^n$ is the polylogarithm function. Substituting \Eq{C(Lambda)} and \Eq{F0} into \Eq{T0,first}, we get
\begin{widetext}
\begin{align}\label{T0}
    T^{(0)}=~&\ln^3\widetilde{B}+\ln(\cos\alpha_1\cos\alpha_2\cos\alpha_3)\ln^2\widetilde{B}+\left[\frac{\pi^2}{4}-\sum_{i=1}^3 \frac{\mathrm{Li}_2(\sin^2\alpha_i)}{2}-6\omega\right]\ln\widetilde{B}
	\nonumber\\
	~&+\left\{\sum_{i=1}^3\left[-2\omega\ln\cos\alpha_i-\frac{\left(\ln\cos\alpha_i\right) \mathrm{Li}_2\left(\cos^2\alpha_i\right)}{2}+\frac{\mathrm{Li}_3\left(\cos^2\alpha_i\right)}{2}\right]\right\}-\frac{1}{2}\ln\left(\frac{4}{3}\right)\mathrm{Li}_2\left(\frac{3}{4}\right)-\mathrm{Li}_3\left(\frac{3}{4}\right),
\end{align}
\end{widetext}
where $\widetilde{B}=\frac{8B}{3\sqrt{3}a}$, and $\alpha_i=\mathrm{arctan}\left(\frac{2R_i}{\sqrt{3}s_i}\right)$
is the hyperangle, satisfying $\sum_{i=1}^3\cos(2\alpha_i)\equiv0$ \cite{tan2008three}.

Expanding $T^{(0)}$ at $R\gg s\gg r_e$, we get
\begin{subequations}
\begin{equation}
t^{(-1,1)}=0,
\end{equation}
    \begin{align}\label{t(-2,2),1st}
        t^{(-2,2)}=~&\frac{s^2}{R^2}\Bigg\{-\frac{3}{8}\left(\ln\frac{s}{a}-1\right)+\Bigg[-\frac{1}{4}\ln^2\widetilde{R}\nonumber\\
        &+\frac{3}{4}\left(-1+3\ln\frac{4}{3}\right)\ln\widetilde{R}+\frac{3}{8}\ln\frac{4}{3}\Bigg]\cos2\theta\Bigg\},
    \end{align}
\end{subequations}
where $\theta=\arccos(\hat{\vc{s}}\cdot \hat{\vc{R}})$ is the angle between $\vc{s}$ and $\vc{R}$.

\paragraph*{\textbf{Step 2: Determination of $T^{(n)}$ and $S^{(n)}$ for $-2<n<0$.}} 
The solution to the \Eq{HsS(-1)} with even $\hat{\vect y}$-parity is
\begin{equation}\label{S(-1) 1st}
    S^{(n)}(\vc{R},\vc{s})=F^{(n)}(R)\phi(\vc{s})+\sum_{l=2,4,\cdots}X^{(n)}_l(R)\phi^{(l)}_{\hat{\vc{R}}}(\vc{s}),
\end{equation}
where $-2<n<0$, and $F^{(n)}(R)$ is a function of the magnitude of $\vc{R}$, and it scales as $R^{n}\ln^{j}{R}$ and $j$ is some number, and similarly for $X^{(n)}_l(R)$. 
The higher partial wave $\phi^{(l)}_{\hat{\vc{R}}}(\vc{s})$ contains  a term proportional to $s^{l}$, so if $X^{(n)}_l(R)\neq0$, the expansion of $S^{(n)}(\vc{R},\vc{s})$ at $R\gg s\gg r_e$ would contain a nonzero term $t^{(n,l)}$ with $l\geq 2$, contradicting \Eq{t,i+j>0 or i>0}. Therefore we must have $X^{(n)}_l(R)=0$, and
\begin{equation}\label{S(-1) 2nd}
    S^{(n)}(\vc{R},\vc{s})=F^{(n)}(R)\phi(\vc{s}).
\end{equation}
Expanding the above equation at $R\gg s\gg r_e$, we find
\begin{subequations}
\begin{equation}\label{t(-1,0)}
    t^{(n,0)}=F^{(n)}(R)\ln\frac{s}{a},
\end{equation}
\begin{equation}
    t^{(n,j)}=0,\quad j\neq0,
\end{equation}
\end{subequations}
where $-2<n<0$.

When $R\gg s\gg r_e$, $T^{(n)}$ for $-2<n<0$ can be further expanded as
\begin{align}\label{T(n) double expansion}
	T^{(n)}
	= t^{(n,0)}+O(R^{n-j}s^j)=F^{(n)}(R)\ln\frac{s}{a}+O(R^{n-j}s^j),
\end{align}
where $j$ is a positive number.

According to \Eq{LaplaceT-i}, $(\nabla_{\vect s_i}^2+\frac34\nabla_{\vect R_i}^2)T^{(n)}=0$ if $0<s_i\ll R_i$.
If we analytically continue $T^{(n)}$ to the nonphysical limit $s_i\to0$, 
we conjecture that $T^{(n)}$ will still contain a term logarithmically dependent on $s_i$, and the coefficient is $F^{(n)}(R_i)$. Therefore,
\begin{equation}\label{T(-1), SE}
    \frac{\nabla_1^2+\nabla_2^2+\nabla_3^2}{2}T^{(n)}=2\pi\sum_{i=1}^3 F^{(n)}(R_i)\delta(\vc{s}_i).
\end{equation}
Using Green's function to solve the above equation, we get
\begin{equation}\label{T(n),first}
    T^{(n)}=-\frac{1}{2\pi}\sum_{i=1}^3\int d^2R'\frac{F^{(n)}(R')}{\frac{3}{4}s_i^2+(\vc{R}_i-\vc{R}')^2}.
\end{equation}
Expanding the above formula at $R\gg s\gg a$, we get 
\begin{align}\label{Fslow(n) integral transform}
T^{(n)}=&\int_0^{\infty}F^{(n)}(R')\Bigg[\frac{Z_{\sqrt{3}a/4}(R'-R)}{|R'-R|}\frac{1}{R'+R}\nonumber\\
    &+\frac{2}{\sqrt{R^4+R'^4+R^2R'^2}}\Bigg]R'dR'+F^{(n)}(R)\ln\frac{s}{a}\nonumber\\
    &+O(R^{n-j}s^j).
\end{align}
Comparing this result with \Eq{T(n) double expansion}, we get an integral equation
satisfied by $F^{(n)}(R)$:
\begin{align}\label{Fslow(n) integral equation}
    &\int_0^{\infty}F^{(n)}(R')\Bigg[\frac{Z_{\sqrt{3}a/4}(R'-R)}{|R'-R|}\frac{1}{R'+R}\nonumber\\
    &+\frac{2}{\sqrt{R^4+R'^4+R^2R'^2}}\Bigg]R'dR'=O( R^{n-j}),
\end{align}
where $j>0$.

If $-2<n<0$ and $j$ is a real number, we find that at large $R$,
\begin{align}
    &\int_{R_0}^{\infty}R'^n\ln^j\widetilde{R}'\Bigg[\frac{Z_{\sqrt{3}a/4}(R'-R)}{|R'-R|}\frac{1}{R'+R}\nonumber\\
    &~~~+\frac{2}{\sqrt{R^4+R'^4+R^2R'^2}}\Bigg]R'dR'\nonumber\\
    =& -R^n\ln^j\widetilde{R}\Bigg[\psi\left(\frac{n}{2}\right)+\frac{\pi}{2}\cot\frac{n\,\pi}{2}+\frac{2}{n}\nonumber\\
    &+\frac{\pi}{\sin\frac{n\,\pi}{2}}P_{-\frac{n+2}{2}}\left(\frac{1}{2}\right) +\gamma+\ln\frac{4}{3}-\ln\widetilde{R}\Bigg]+o(R^n\ln^j\widetilde{R}),
\end{align}
where $R_0$ is a fixed length scale satisfying $R_0>3\sqrt3\,a/8$,
$\widetilde{R}'=8R'/(3\sqrt3\,a)$,
 $P_\nu(z)$ is the  Legendre function \cite{Olver2010NIST}, and  $\psi(z)$ is the digamma function defined as the derivative of the natural logarithm of the Gamma function: $\psi(z)\equiv \frac{d}{dz}\ln\Gamma(z)$. For $-2<n<0$,
\begin{align}
    &\psi\left(\frac{n}{2}\right)+\frac{\pi}{2}\cot\frac{n\,\pi}{2}+\frac{2}{n}+\frac{\pi}{\sin\frac{n\,\pi}{2}}P_{-\frac{n+2}{2}}\left(\frac{1}{2}\right)+\gamma +\ln\frac{4}{3}\nonumber\\
    \leq &-2K\left(\frac{1}{4}\right)-\ln 3=-4.47011\ldots,
\end{align}
where $K(z)=\int_0^{\pi/2}\frac{d\theta}{\sqrt{1-z\sin^2\theta}}$ is the complete elliptic integral of the first kind.
Therefore, there is no nonzero solution to \Eq{Fslow(n) integral equation} of order $R^n\ln^j\widetilde{R}$ for $-2<n<0$. It follows that
\begin{equation}
    S^{(n)}=0, ~\text{if }-2<n<0,   
\end{equation}
\begin{equation}
T^{(n)}=0,~\text{if }-2<n<0.
\end{equation}

\paragraph*{\textbf{Step 3: Determination of $T^{(-2)}$ and $S^{(-2)}$.}} 
Using Eqs.~\eqref{HsS(-2)}, \eqref{S0 2nd}, and \eqref{F0}, we get
\begin{equation}\label{HS(-2)}
    H_{\vc{s}}S^{(-2)}(\vc{R},\vc{s})
    =\frac{3}{2R^2} \phi(s).
\end{equation}
Solving \Eq{HS(-2)} and using Eqs.~(\ref{Hphi2body}), we get
\begin{align}\label{S(-2)}
    S^{(-2)}=~&F^{(-2)}(R)\phi(\vc{s})+\frac{3}{2R^2} f(\vc{s})+H^{(-2)}(R)\phi^{(d)}_{\hat{\vc{R}}}(\vc{s})\nonumber\\
    &+\sum_{l=4,6,\cdots}X^{(-2)}_l(R)\phi^{(l)}_{\hat{\vc{R}}}(\vc{s}),
\end{align}
where $F^{(-2)}(R), H^{(-2)}(R), X^{(-2)}_l(R)\sim R^{-2}\ln^n R$. The higher partial wave $\phi^{(l)}_{\hat{\vc{R}}}(\vc{s})$ contains  a term proportional to $s^{l}$ at $s>r_e$, so if $X^{(-2)}_l(R)\neq0$ at $l\ge4$, the expansion of $S^{(0)}(\vc{R},\vc{s})$ at $R\gg s\gg r_e$ would contain a nonzero $t^{(-2,l)}$($l\geq 4$), contradicting \Eq{t,i+j>0 or i>0}. Thus, $X^{(-2)}_l(R)=0$ for $l\ge4$, and
\beq
S^{(-2)}=F^{(-2)}(R)\phi(\vc{s})+\frac{3}{2R^2} f(\vc{s})+H^{(-2)}(R)\phi^{(d)}_{\hat{\vc{R}}}(\vc{s}).
\eeq
Expanding $S^{(-2)}$ at $R\gg s\gg r_e$, we get
\begin{equation}
    S^{(-2)}=t^{(-2,2)}+t^{(-2,0)}+t^{(-2,-2)}+O(R^{-2}s^{-4}),
\end{equation}
where
\begin{subequations}
\begin{align}
    t^{(-2,2)}=~& -\frac{3}{8R^2}s^2\left(\ln\frac{s}{a}-1\right)+H^{(-2)}(R)\frac{s^2}{8}\cos2\theta,\label{t(-2,2),2nd}\\
    t^{(-2,0)}=&-\frac{3\pi r_s}{8R^2}+F^{(-2)}(R)\ln\frac{s}{a},\label{t(-2,0)}\\
    t^{(-2,-2)}=~& H^{(-2)}(R)\left(-\frac{4a_d}{\pi s^2}\right)\cos2\theta,\label{t(-2,-2),1st}
\end{align}
\end{subequations}
and $\theta$ is the angle between $\hat{\vect s}$ and $\hat{\vect R}$.
Comparing \Eq{t(-2,2),1st} with \Eq{t(-2,2),2nd}, we get
\begin{equation}\label{H(-2)}
    H^{(-2)}(R)=-\frac{2}{R^2}\left[\ln^2{\widetilde{R}}+3\left(1-3\ln\frac{4}{3}\right)\ln{\widetilde{R}}-\frac{3}{2}\ln\frac{4}{3}\right].
\end{equation}
According to \Eq{LaplaceT-i}, if $s_1, s_2, s_3>0$,
\begin{align}\label{T(-2) SE}
    \frac{\nabla_1^2+\nabla_2^2+\nabla_3^2}{2}T^{(-2)}=0.
\end{align}
When $R\gg s\gg r_e$,
\begin{align}\label{T(-2) expansion}
    T^{(-2)}&=t^{(-2,0)}++O(R^{-2-j}s^j)\nonumber\\
    &=-\frac{3\pi r_s}{8R^2}+F^{(-2)}(R)\ln\frac{s}{a}+O(R^{-2-j}s^j),
\end{align}
where $j$ is a number satisfying $j>0$.
If we analytically continue $T^{(-2)}$ to the nonphysical limit $s\to0$, 
we conjecture that $T^{(-2)}$ will still contain a term logarithmically dependent on $s$, and the coefficient is still $F^{(-2)}(R)$.
Therefore,
\begin{align}\label{T(-2) SE2}
    &\frac{\nabla_1^2+\nabla_2^2+\nabla_3^2}{2}T^{(-2)}\nonumber\\
    =~&2\pi\sum_{i=1}^3 \left[F^{(-2)}(R_i)+c\, \delta(\vc{R}_i)\right]\delta(\vc{s}_i),
\end{align}
where $c$ is some number. Without loss of generality, we assume that
\begin{equation}\label{F-2wholespace}
    \int F^{(-2)}(R') d^2R'=0.
\end{equation}
If this is not the case, for example if $\int F^{(-2)}(R') d^2R'=\tau$, we can define
\begin{equation}
    F'^{(-2)}(R)=F^{(-2)}(R)-\tau\,\delta(\vc{R}),
\end{equation}
and
\begin{equation}
    c'=c+\tau,
\end{equation}
such that
\begin{equation}
    \int F'^{(-2)}(R') d^2R'=0,
\end{equation}
\begin{align}
    &\frac{\nabla_1^2+\nabla_2^2+\nabla_3^2}{2}T^{(-2)}\nonumber\\
    =~&2\pi\sum_{i=1}^3 \left[F'^{(-2)}(R_i)+c'\, \delta(\vc{R}_i)\right]\delta(\vc{s}_i).
\end{align}
The solution to \Eq{T(-2) SE2} is
\begin{equation}\label{T(-2),1st}
    T^{(-2)}=-\frac{1}{2\pi}\sum_{i=1}^3\int d^2R'\frac{F^{(-2)}(R')+c\, \delta(\vc{R}')}{\frac{3}{4}s_i^2+(\vc{R}_i-\vc{R}')^2}.
\end{equation}
Expanding the above formula at $R\gg s\gg a$ and comparing the result with \Eq{T(-2) expansion}, 
we find 
\begin{equation}\label{c}
    c=\frac{\pi^2 r_s}{4},
\end{equation}
\begin{align}\label{F(-2) equ 1st}
    &\int_0^{\infty}F^{(-2)}(R')\Bigg[\frac{Z_{\sqrt{3}a/4}(R'-R)}{|R'-R|}\frac{1}{R'+R}\nonumber\\
    &+\frac{2}{\sqrt{R^4+R'^4+R^2R'^2}}\Bigg]R'dR'=O(R^{-2-j}),
\end{align}
where $j>0$.
We try the following solution at $R\to\infty$:
\beq\label{F-2approx}
F^{(-2)}(R)\approx-\frac{D}{4\pi^2R^2\ln^n\widetilde{R}},
\eeq
where $D$ is some constant dependent on the interactions, and $n>1$ in order to be consistent with \Eq{F-2wholespace}.
We define a generalized function $\frac{Z}{R^2\ln^n\widetilde{R}}$ on the $\vect R$ plane satisfying
\begin{subequations}
\begin{equation}
    \frac{Z}{R^2\ln^n\widetilde{R}}=\frac{1}{R^2\ln^n\widetilde{R}},~\widetilde{R}>1,
\end{equation}
\begin{equation}
    \int\frac{Z}{R^2\ln^n\widetilde{R}}d^2R=0.
\end{equation}
\end{subequations}
When $R\to\infty$,
\begin{align}\label{ZlnR integral transform}
    &\int_0^{\infty}\frac{Z}{R'^2\ln^n\widetilde{R}'}\Bigg[\frac{Z_{\sqrt{3}a/4}(R'-R)}{|R'-R|}\frac{1}{R'+R}\nonumber\\
    &~~~+\frac{2}{\sqrt{R^4+R'^4+R^2R'^2}}\Bigg]R'dR'\nonumber\\
    =&\frac{n-4}{n-1}\frac{1}{R^2\ln^{n-1}\widetilde{R}}+\sum_{j=1}^{J}\frac{K_{n,n+j}}{R^2\ln^{n+j}\widetilde{R}}\nonumber\\
    &+O(R^{-2}\ln^{-n-J-1}\widetilde{R})
\end{align}
for any positive integer $J$, where
\begin{align}
    K_{n,n+j}=~&\binom{-n}{j}\Bigg\{\left[\left(\frac{1}{2}\right)^{j+1}-\left(-\frac{1}{2}\right)^{j+1}\right]j!\zeta(j+1)\nonumber\\
    &+2(-1)^j B_j\Bigg\},
\end{align}
where $\binom{-n}{j}$ is the binomial coefficient: $\binom{-n}{j}\equiv (-1)^j\frac{(n+j-1)!}{j!(n-1)!}$,
and
\begin{align}
    B_j=~&\lim_{\Lambda\to\infty}\int_0^\Lambda\frac{x\ln^j x}{\sqrt{1+x^2+x^4}}dx-\frac{\ln^{j+1}\Lambda}{j+1}\nonumber\\
    =~&\lim_{\xi\to2}\frac{d^j}{d\xi^j}\left[\frac{\pi P_{-\frac{\xi}{2}}\left(\frac{1}{2}\right)}{2\sin\frac{\pi\xi}{2}}+\frac{1}{\xi-2}\right]\nonumber\\
    =~& j!\sum_{n=1}^{\infty}(-1)^n\left[\frac{ P_{-n-1}\left(\frac{1}{2}\right)}{(2n)^{j+1}}+\frac{ P_{n-1}\left(\frac{1}{2}\right)}{(-2n)^{j+1}}\right].
\end{align}
Substituting \Eq{F-2approx} into \Eq{F(-2) equ 1st}, and using \Eq{F-2wholespace} and \Eq{ZlnR integral transform}, we find that $n=4$ in \Eq{F-2approx}.
Equation~\eqref{ZlnR integral transform} implies a more accurate formula for $F^{(-2)}(R)$ at $R\to\infty$:
\begin{equation}\label{Fslow series}
    F^{(-2)}(R)=-\frac{D}{4\pi^2}\sum_{n=4}^{n_\text{max}}\frac{d_n}{R^2\ln^n\widetilde{R}}+O(R^{-2}\ln^{-n_\text{max}-1}\widetilde{R}),
\end{equation}
where $n_\text{max}$ is an arbitrary number equal to or greater than 4, and
\begin{subequations}
\begin{equation}
    d_4\equiv1.
\end{equation}
$d_n$ ($n\ge5$) are some constants to be determined. From Eqs.~\eqref{F(-2) equ 1st}, \eqref{ZlnR integral transform}, and \eqref{Fslow series}, we derive
\begin{equation}
    d_5=0
\end{equation}
and a recurrence relation for $d_6$, $d_7$, $d_8$, \dots:
\begin{align}
    d_{n+1}=-\frac{\sum_{l=4}^{n-1}K_{l,n}d_l}{1-\frac{3}{n}},~\text{if }n \geq 5.
\end{align}
\end{subequations}
The analytical expressions of $d_6$ and $d_7$ are
\begin{subequations}\label{d567}
    \begin{equation}
        d_6=20\omega,
    \end{equation}
    \begin{align}
        d_7=~&10\left[\ln\left(\frac{4}{3}\right)\mathrm{Li}_2\left(\frac{3}{4}\right)+2\mathrm{Li}_3\left(\frac{3}{4}\right)-3\zeta(3)\right],
    \end{align}
\end{subequations}
where $\omega$ is defined in \Eq{omega}. The numerical values of $d_4$ through $d_{20}$ are listed in Table~\ref{table:d_n}.

\begin{table}
\centering
\sisetup{table-format=13.9, table-number-alignment=center} 
\begin{tabular}{ |c|S| } 
\hline
$d_4$ & 1 \\
\hline
$d_5$ & 0 \\
\hline
$d_6$ & 6.886407139 \\
\hline
$d_7$ & -16.358309890 \\
\hline
$d_8$ & 88.433470958 \\
\hline
$d_9$ & -442.993797988 \\
\hline
$d_{10}$ & 2401.809713951 \\
\hline
$d_{11}$ & -14662.976771313 \\
\hline
$d_{12}$ & 95130.449443130 \\
\hline
$d_{13}$ & -668591.531772193 \\
\hline
$d_{14}$ & 5023390.160301575 \\
\hline
$d_{15}$ & -40254096.130488401 \\
\hline
$d_{16}$ & 342633436.373693474 \\
\hline
$d_{17}$ & -3087180147.856111754 \\
\hline
$d_{18}$ & 29356322170.257811613 \\
\hline
$d_{19}$ & -293801482972.836230509 \\
\hline
$d_{20}$ & 3087060711505.942650267 \\
\hline
\end{tabular}
\caption{\label{table:d_n}The numerical values of $d_n$ for $4\le n\le 20$.}
\end{table}

\begin{figure}
    \centering
    \includegraphics[width=0.45\textwidth]{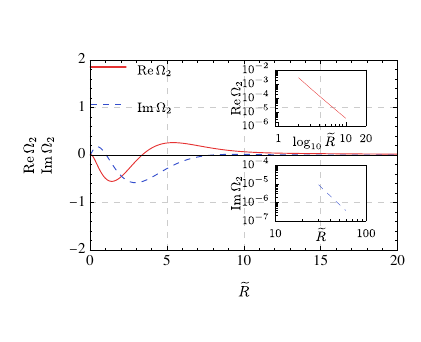}
    \caption{
    Real (red solid line) and imaginary (blue dashed line) parts of $\Omega_2$ as functions of $\widetilde{R}$. Insets show the real and the imaginary parts of $\Omega_2$ at large values of $\log_{10}\widetilde{R}$ or $\widetilde{R}$, and they are consistent with \Eq{Omega2 expansion} and \Eq{ImOmega2} respectively.      }
    \label{fig:Omega2&4}
\end{figure}

Clearly $\frac{d_{n+1}}{d_n}\to-\infty$ at $n\to\infty$.
So the right hand side of \Eq{Fslow series} is an asymptotic series whose radius of convergence is zero.
In order to derive a more accurate formula for $F^{(-2)}(R)$, we look for an auxiliary function $F^{(-2)}_\text{aux}$ satisfying
\begin{align}\label{Faux integral equation}
    &\int_0^{\infty}F^{(-2)}_\text{aux}(R')\Bigg[\frac{Z_{\sqrt{3}a/4}(R'-R)}{|R'-R|}\frac{1}{R'+R}\nonumber\\
    &+\frac{2}{\sqrt{R^4+R'^4+R^2R'^2}}\Bigg]R'dR'=0.
\end{align}
We write a trial solution
\begin{equation}\label{Faux}
F^{(-2)}_\text{aux}(R)=\frac{1}{2\pi\mathrm{i}\,R^2}\lim_{\epsilon\to0^+}\int_{0}^{\mathrm{i}\infty}\widetilde{R}^{\xi}M(\xi)e^{-\epsilon|\xi|}d\xi,
\end{equation}
where $M(\xi)$ is a function to be determined.
Substituting \Eq{Faux} into \Eq{Faux integral equation}, we find that $M(\xi)$ satisfies a differential equation:
\begin{align}
&\frac{dM(\xi)}{d\xi}+\Bigg[-\frac14\psi\Big(\frac{1-\xi}{2}\Big)+\frac14\psi\Big(1-\frac{\xi}{2}\Big)+\frac12\psi(1-\xi)\nonumber\\
&+\frac12\psi\Big(\frac\xi2\Big)-\ln\frac{3\sqrt2\,e^{-\gamma}}{4}-\frac{\pi P_{-\frac{\xi}{2}}\left(\frac{1}{2}\right)}{\sin\frac{\pi \xi}{2}}\Bigg]M(\xi)=0,
\end{align}
which can be written as
\begin{align}\label{M(xi) equ}
    &\frac{d}{d\xi} \ln \left\{M(\xi)/\left[\left(\frac{3e^{-\gamma}}{4}\right)^{\xi}\Big(\sin\frac{\pi\xi}{2}\Big)\Gamma^2\left(1-\frac{\xi}{2}\right)\right]\right\}\nonumber\\
    &=\frac{\pi P_{-\frac{\xi}{2}}\left(\frac{1}{2}\right)}{\sin\frac{\pi \xi}{2}}.
\end{align}
Solving the above equation, we get
\begin{equation}\label{M(xi)}
    M(\xi)=c_M\left(\frac{3e^{-\gamma}}{4}\right)^{\xi}\Big(\sin\frac{\pi\xi}{2}\Big)\Gamma^2\left(1-\frac{\xi}{2}\right)L(\xi),
\end{equation}
where $c_M$ is a constant, and
\begin{equation}
    L(\xi)=\exp\left[\int_{\mathrm{Re}(\xi)+\mathrm{i}\infty}^{\xi}\frac{\pi P_{-\frac{t}{2}}\left(\frac{1}{2}\right)}{\sin\frac{\pi t}{2}}dt\right],
\end{equation}
where $\mathrm{Re}$ stands for the real part.
When $\xi\to0$,
\beq\label{M small xi}
M(\xi)=-\frac{\pi \beta c_M}{2}\xi^3+O(\xi^4),
\eeq
where
\begin{align}
\beta&=\exp\bigg[\lim_{\epsilon\to0^+}-2\ln\epsilon+\int_{\I\infty}^{\I\epsilon}\frac{\pi P_{-\frac{t}{2}}(\frac12)}{\sin\frac{\pi t}{2}}dt\bigg]\nonumber\\
&\approx0.425611- 0.224837\,\mathrm{i}.
\end{align}
Substituting \Eq{M small xi} into \Eq{Faux}, we find that at $\widetilde R\to\infty$ the function $F^{(-2)}_\text{aux}(R)$ contains
a term equal to $3\I\beta c_M/[2R^2\ln^4\widetilde{R}]$. We choose $c_M$ such that this term is simply $1/[R^2\ln^4\widetilde{R}]$.
Thus
\beq
c_M=-\frac{2\mathrm{i}}{3\beta}.
\eeq
From \Eq{Faux} and \Eq{M(xi)}, we derive that
\begin{equation}\label{Faux 2D integral}
    \int F^{(-2)}_\text{aux}(R')d^2R'=0,
\end{equation}
and that $F^{(-2)}_\text{aux}(R)$ contains an oscillating term $-\frac{\mathrm{i}\pi b_0^2}{6\beta}H_0^{(1)}(b_0 R)$ at large $R$, where 
\beq
b_0=\frac{4e^{-\gamma}}{\sqrt3\,a},
\eeq
and $H_l^{(1)}(z)$ is the Hankel function of the first kind \cite{morse1953methods}.
If this oscillating term is subtracted from $F^{(-2)}_\text{aux}(R)$, 
we find that when $\widetilde{R}\gg 1$, the remaining part has the same asymptotic expansion as
the asymptotic series on the right hand side of \Eq{Fslow series}:
\begin{align}
    &F^{(-2)}_\text{aux}(R)+\frac{\mathrm{i}\pi b_0^2}{6\beta}H_0^{(1)}(b_0 R)\nonumber\\
    &=\sum_{n=4}^{n_\text{max}}\frac{d_n}{R^2\ln^n\widetilde{R}}+O(R^{-2}\ln^{-n_\text{max}-1}\widetilde{R}).
\end{align}
On the other hand, at large $R$,
\begin{align}\label{hankel delta integral transform}
    &\int_0^{\infty}\left[H_0^{(1)}(b_0 R')-\frac{4\,\mathrm{i}}{b_0^2}\frac{\tilde{\delta}(R')}{2\pi }\right]\nonumber\\
    &\Bigg[\frac{Z_{\sqrt{3}a/4}(R'-R)}{|R'-R|}\frac{1}{R'+R}+\frac{2}{\sqrt{R^4+R'^4+R^2R'^2}}\Bigg]R'dR'\nonumber\\
    &=O( R^{-6}),
\end{align}
where $\tilde{\delta}(R)$ is a generalized function defined such that
\begin{subequations}
    \begin{equation}
        \tilde{\delta}(R)=0,~\text{if }R>0,
    \end{equation}
    \begin{equation}
        \int_0^\infty \tilde{\delta}(R)R dR=1.
    \end{equation}
\end{subequations}
From Eqs.~(\ref{F(-2) equ 1st}), (\ref{Fslow series}), (\ref{Faux integral equation}), (\ref{Faux 2D integral})--(\ref{hankel delta integral transform}), we get
\begin{align}\label{F(-2) 1st}
    F^{(-2)}(R)=&-\frac{D}{4\pi^2}\Big[F^{(-2)}_\text{aux}(R)+\frac{\mathrm{i}\pi b_0^2}{6\beta}H_0^{(1)}(b_0 R)\nonumber\\
    &\mspace{60mu}+\frac{1}{3\beta}\tilde{\delta}(R)+\widetilde{F}^{(-2)}_\text{aux}(R)\Big],
\end{align}
where $\widetilde{F}^{(-2)}_\text{aux}(R)$ is a function satisfying
\beq
\int \widetilde{F}^{(-2)}_\text{aux}(R)d^2R=0
\eeq
and
\beq
 \widetilde{F}^{(-2)}_\text{aux}(R)=O(R^{-2-j}),~R\to\infty,
\eeq
for some $j>0$.
We introduce a mathematical function $\Omega_2(\widetilde{R})$  
\begin{align}\label{Omega2 analytic}
    \Omega_2(\widetilde{R})&=R^2\left[F^{(-2)}_\text{aux}(R)+\frac{\mathrm{i}\pi b_0^2}{6\beta}H_0^{(1)}(b_0 R)\right]\nonumber\\
    &=\frac{1}{2\pi\mathrm{i}}\lim_{\epsilon\to0^+}\int_{0}^{\mathrm{i}\infty}\widetilde{R}^{\xi}M(\xi)e^{-\epsilon|\xi|}d\xi\nonumber\\
    &\quad+\frac{\mathrm{i}\pi}{6\beta}\left(\frac{3e^{-\gamma}}{2}\widetilde{R}\right)^2H_0^{(1)}\left(\frac{3e^{-\gamma}}{2}\widetilde{R}\right)\nonumber\\
    &=-\frac{1}{6\pi\beta}\int^{\mathrm{i}\infty}_{0}\left(\frac{3e^{-\gamma}}{4}\widetilde{R}\right)^\xi \Gamma^2\left(1-\frac{\xi}{2}\right)\nonumber\\
    &\mspace{120mu}\Big[2L(\xi)\sin\frac{\pi\xi}{2}-\mathrm{i}e^{-\mathrm{i}\frac{\pi}{2}\xi}\Big]d\xi\nonumber\\
    &\quad+\frac{\mathrm{i}}{6\pi\beta}\int_{-\mathrm{i}\infty}^{0}\left(\frac{3e^{-\gamma}}{4}\widetilde{R}\right)^\xi e^{-\mathrm{i}\frac{\pi}{2}\xi}\Gamma^2\left(1-\frac{\xi}{2}\right)d\xi.
\end{align}
We can reexpress $F^{(-2)}(R)$ in \Eq{F(-2) 1st} in terms of $\Omega_2(\widetilde{R})$:
\beq
F^{(-2)}(R)=-\frac{D}{4\pi^2}\Big[\frac{\Omega_2(\widetilde{R})}{R^2}+\frac{1}{3\beta}\tilde{\delta}(R)+\widetilde{F}^{(-2)}_\text{aux}(R)\Big].
\eeq
When $\widetilde{R}\gg 1$,
\begin{equation}\label{Omega2 expansion}
    \mathrm{Re}\,\Omega_2(\widetilde{R})=\sum_{n=4}^{n_\text{max}}\frac{d_n}{\ln^n\widetilde{R}}+O(\ln^{-n_\text{max}-1}\widetilde{R}),
\end{equation}
and
\beq\label{ImOmega2}
\mathrm{Im}\,\Omega_2(\widetilde{R})=\omega_2\widetilde{R}^{-4}\ln^{-\frac{7}{4}}\widetilde{R}+O\left(\widetilde{R}^{-4}\ln^{-\frac{11}{4}}\widetilde{R}\right),
\eeq
where $\mathrm{Im}$ stands for the imaginary part, and
\begin{align}
\omega_2&=\frac{128\pi e^{4\gamma}}{81\Gamma(\frac14)}\exp\bigg[\lim_{\epsilon\to0^+}
\frac94\ln\epsilon-P\int_{-4+\epsilon}^{-\epsilon}\frac{\pi P_{-\frac{t}{2}}(\frac12)}{\sin\frac{\pi t}{2}}dt\bigg]\nonumber\\
&\approx~35.7044,
\end{align}
where $P\int$ is the principal value integral for the pole at $t=-2$.
In Fig.~\ref{fig:Omega2&4} we plot the function $\Omega_2(\widetilde{R})$.
Since the imaginary part of $\frac{\Omega_2(\widetilde{R})}{R^2}$ decays like $R^{-6}\ln^{-\frac{7}{4}}\widetilde{R}$ at $R\to+\infty$, we may choose 
$\widetilde{F}^{(-2)}_\text{aux}(R)=-\mathrm{i}\,\mathrm{Im}\left[\frac{\Omega_2(\widetilde{R})}{R^2}+\frac{1}{3\beta}\tilde{\delta}(R)\right]$, and write
\beq\label{F(-2)}
F^{(-2)}(R)=-\frac{D}{4\pi^2}\Big[\frac{\mathrm{Re}\,\Omega_2(\widetilde{R})}{R^2}+\frac{\mathrm{Re}\,\beta}{3|\beta|^2}\widetilde{\delta}(R)\Big].
\eeq

\begin{figure*}
    \centering
    \includegraphics[width=\textwidth]{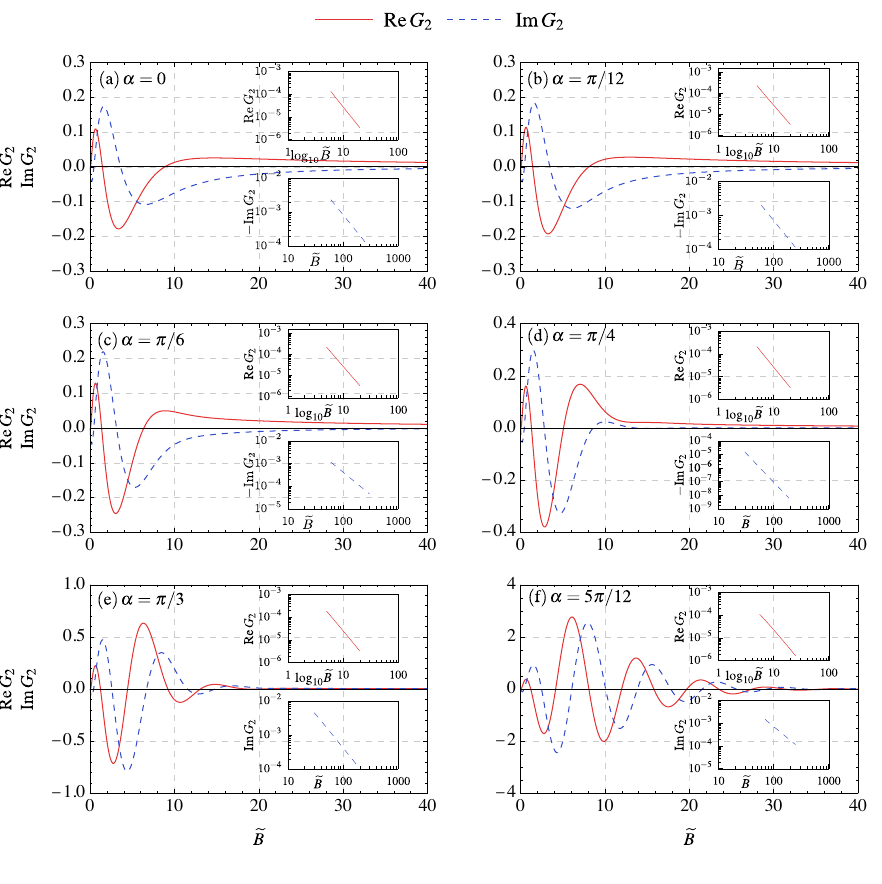}
    \caption{The real part (red solid line) and the imaginary part (blue dashed line) of $G_2(\widetilde{B},\alpha)$ as functions of $\widetilde{B}$ 
        for hyper-angles $\alpha = 0$, $\pi/12$, $\pi/6$, $\pi/4$, $\pi/3$, $5\pi/12$. 
        The insets show the real and the imaginary parts of $G_2(\widetilde{B},\alpha)$ at large $\widetilde{B}$. 
        }
    \label{fig:G2}
\end{figure*}

Substituting \Eq{c} and \Eq{F(-2)} into \Eq{T(-2),1st}, and using \Eq{Omega2 analytic}, we get
\begin{equation}\label{T(-2)}
    T^{(-2)}=-\frac{3\pi\, r_s}{8 B^2} -\frac{D}{4\pi^2 B^2}\sum_{i=1}^3 \mathrm{Re}\,G_2(\widetilde{B},\alpha_i)-\frac{D}{4\pi^2}\sum_{i=1}^3 \widetilde{T}_i^{(-2)},
\end{equation}
where $G_2(\widetilde{B},\alpha)$ is a mathematical function defined as
\begin{equation}\label{G2 definition}
    G_2(\widetilde{B},\alpha)\equiv-\frac{B^2}{2\pi}\int d^2R'\frac{F^{(-2)}_\text{aux}(R')}{\frac{3}{4}s^2+(\vc{R}-\vc{R}')^2},
\end{equation}
where $B=\sqrt{R^2+3s^2/4}$, $\widetilde{B}=\frac{8B}{3\sqrt{3}a}$, $\alpha=\arctan(2R/\sqrt3\,s)$, and
\begin{align}
    \widetilde{T}_i^{(-2)}=~&-\frac{1}{2\pi}\sum_{i=1}^3\mathrm{Re}\int d^2R'\frac{\frac{\mathrm{i}\pi b_0^2}{6\beta}H_0^{(1)}(b_0 R')+\frac{1}{3\beta}\tilde{\delta}(R')}{\frac{3}{4}s_i^2+(\vc{R}_i-\vc{R}')^2}\nonumber\\
    =~&O(B^{-4}),~~B\to\infty.
\end{align}
At $B\to\infty$, we thus have 
\begin{equation}\label{T(-2), 2nd}
    T^{(-2)}=-\frac{3\pi\, r_s}{8 B^2} -\frac{D}{4\pi^2 B^2}\sum_{i=1}^3 \mathrm{Re}\,G_2(\widetilde{B},\alpha_i)+O(B^{-4}).
\end{equation}
Substituting \Eq{Faux} into \Eq{G2 definition}, we get
\begin{align}\label{G2 analytic}
    G_2(\widetilde{B},\alpha)=~&-\frac{1}{2\pi\mathrm{i}}\int_{0}^{\mathrm{i}\infty}\widetilde{B}^{\xi}M(\xi)\frac{\pi P_{-\frac{\xi}{2}}\left(\cos2\alpha\right)}{2\sin\frac{\pi\xi}{2}}d\xi.
\end{align}
At large $\widetilde{B}$, $\mathrm{Re}\,G_2(\widetilde{B},\alpha)$ has the following asymptotic series expansion in powers of $1/\ln\widetilde{B}$:
\begin{equation}\label{G2 expansion}
    \mathrm{Re}\,G_2(\widetilde{B},\alpha)=\sum_{n=3}^{n_\text{max}} \frac{h_n(\alpha)}{\ln^n\widetilde{B}}+O\left(\frac{1}{\ln^{n_\text{max}+1}\widetilde{B}}\right),
\end{equation}
where
\begin{align}\label{hn}
    h_n(\alpha)=\frac{d_{n+1}}{n}-\sum_{l=0}^{n-4}\binom{l-n}{l}d_{n-l}2^{-l-1}W_l(\alpha).
\end{align}
The function $W_l(\alpha)$ in \Eq{hn} is defined as
\begin{align}\label{Wl}
    W_l(\alpha)=~&\lim_{\lambda\to0}\int_\lambda^\infty\frac{\ln^l x}{x\sqrt{1+2\left(\cos2\alpha\right)x+x^2}}dx\nonumber\\
    &+\frac{\ln^{l+1}\lambda}{l+1}\nonumber\\
    =~&\lim_{\xi\to0}\frac{d^l}{d\xi^l}\left[ \frac{\pi P_{-\frac{\xi}{2}}\left(\cos 2\alpha\right)}{\sin\frac{\pi\xi}{2}}-\frac{2}{\xi}\right]2^l.
\end{align}
In particular,
\begin{subequations}\label{W012}
    \begin{equation}
        W_0(\alpha)=-2\ln\cos\alpha,
    \end{equation}
    \begin{equation}
        W_1(\alpha)=\frac{\pi^2}{6}-\mathrm{Li}_2\left(\sin^2\alpha\right),
    \end{equation}
    \begin{equation}
        W_2(\alpha)=4\left[\ln\left(\cos\alpha\right)\mathrm{Li}_2\left(\cos^2\alpha\right)-\mathrm{Li}_3\left(\cos^2\alpha\right)+\zeta(3)\right].
    \end{equation}
\end{subequations}
The expansion in \Eq{G2 expansion}, up to the order $\frac{1}{\ln^6\widetilde{B}}$, is
\begin{align}
&\mathrm{Re}\,G_2(\widetilde{B},\alpha)=\frac{1}{3\ln^3\widetilde{B}}+\frac{\ln\cos\alpha}{\ln^4\widetilde{B}}+\frac{4\,\omega+\frac{\pi^2}{6}-\mathrm{Li}_2\left(\sin^2\alpha\right)}{\ln^5\widetilde{B}}\nonumber\\
	&+\Bigg[\frac{5}{3}\ln\left(\frac{4}{3}\right)\mathrm{Li}_2\left(\frac{3}{4}\right)+\frac{10}{3}\mathrm{Li}_3\left(\frac{3}{4}\right)-10\,\zeta(3)\nonumber\\
 &+20\,\omega\ln\left(\cos\alpha\right)+5\ln\left(\cos\alpha\right)\mathrm{Li}_2\left(\cos^2\alpha\right)\nonumber\\
 &+5\,\mathrm{Li}_3\left(\cos^2\alpha\right)\Bigg]/{\ln^6\widetilde{B}}+O\left(\ln^{-7} \widetilde{B}\right).
\end{align}
At large $\widetilde{B}$, 
\begin{equation}\label{Im G2 expansion}
    \mathrm{Im}\, G_2(\widetilde{B},\alpha)=\omega_2'\frac{\cos 2\alpha}{\widetilde{B}^2}+O\left(\widetilde{B}^{-4}\ln^{-\frac{3}{4}}\widetilde{B}\right),
\end{equation}
where
\begin{align}
    \omega_2'&=-\frac{8\pi e^{2\gamma}}{27}\exp\bigg[\lim_{\epsilon\to0^+}
3\ln\epsilon-\int_{-2+\epsilon}^{-\epsilon}\frac{\pi P_{-\frac{t}{2}}(\frac12)}{\sin\frac{\pi t}{2}}dt\bigg]\nonumber\\
&\approx-8.28384.
\end{align}
Fig.~\ref{fig:G2} plots $G_2(\widetilde{B},\alpha)$ versus $\widetilde{B}$  at fixed hyper-angles $\alpha = 0$, $\pi/12$, $\pi/6$, $\pi/4$, $\pi/3$, $5\pi/12$, along with the asymptotic behaviors of the real and the imaginary parts of $G_2(\widetilde{B},\alpha)$ at large $\widetilde{B}$. The asymptotic behaviors of $\mathrm{Re}\,G_2(\widetilde{B},\alpha)$ and $\mathrm{Im}\,G_2(\widetilde{B},\alpha)$ at large $\widetilde{B}$ are consistent with \Eq{G2 expansion} and \Eq{Im G2 expansion}, respectively.

\paragraph*{\textbf{Summary.}} 
The above step-by-step procedure can be continued, but we stop at $S^{(-2)}$  and $T^{(-2)}$, and sum up all $S^{(n)}$ ($0\ge n\ge -2$) we have derived to get the ``21-expansion", and sum up all $T^{(n)}$ ($0\ge n\ge -2$) we have derived to get the ``111-expansion". 

When the pairwise distances
$s_1$, $s_2$, and $s_3$ go to infinity simultaneously,
the wave function has the ``111-expansion":
\begin{widetext}
\begin{subequations}
\begin{align}\label{111expansion}
	\phi^{(3)}(\vc{r}_1,\vc{r}_2,\vc{r}_3)=~&\ln^3\widetilde{B}+\ln(\cos\alpha_1\cos\alpha_2\cos\alpha_3)\ln^2\widetilde{B}+\left[\frac{\pi^2}{4}-\sum_{i=1}^3 \frac{\mathrm{Li}_2(\sin^2\alpha_i)}{2}-6\omega\right]\ln\widetilde{B}
	\nonumber\\
	~&+\Bigg\{\sum_{i=1}^3\left[-2\omega\ln\cos\alpha_i-\frac{\left(\ln\cos\alpha_i\right) \mathrm{Li}_2\left(\cos^2\alpha_i\right)}{2}+\frac{\mathrm{Li}_3\left(\cos^2\alpha_i\right)}{2}\right]\Bigg\}-\frac{1}{2}\ln\left(\frac{4}{3}\right)\mathrm{Li}_2\left(\frac{3}{4}\right)-\mathrm{Li}_3\left(\frac{3}{4}\right)\nonumber\\
	&-\frac{3\pi r_s}{8 B^2} -\frac{D}{4\pi^2 B^2}\sum_{i=1}^3\mathrm{Re}\,G_2(\widetilde{B},\alpha_i)+O(B^{-2-j}),
\end{align}
where $j$ is some positive number.

When $\vect s$ is fixed, but $R\to\infty$,
the wave function has the ``21-expansion":
\begin{align}\label{21expansionrealspace}
	\phi^{(3)}(\vc{R},\vc{s}/2,-\vc{s}/2)=&\left[\ln^{2}{\widetilde{R}}-2\omega-\frac{D}{4\pi^2 R^2}\mathrm{Re}\,\Omega_2(\widetilde{R})\right]\phi(\vc{s})+\frac{3}{2R ^2}f(\vc{s})\nonumber\\
 &-\frac{2}{R^2}\left[\ln^2{\widetilde{R}}+3\left(1-3\ln{\frac{4}{3}}\right)\ln{\widetilde{R}}-\frac{3}{2}\ln{\frac{4}{3}}\right]\phi_{\hat{\vc{R}}}^{(d)}(\vc{s})+O(R^{-2-j'}),
\end{align}
\end{subequations}
\end{widetext}
where $j'$ is some positive number.

If there are two-body bound states, the three incoming free bosons with zero total energy
may recombine into a dimer and a free boson, which fly apart with total kinetic energy equal to the released two-body binding energy, and the 21-expansion is modified as \cite{zhu2017three}
\begin{widetext}
\begin{equation}\label{21expansionnew}
    \phi^{(3)}(\vc{s}/2,-\vc{s}/2,\vc{R})=\Phi_{21}+\sum_{l=0,2,4,\cdots,l_{\text{max}}}\sum_{\nu=0}^{\nu_l}C_{l\nu}\varphi_{l\nu}(\vc{s},\vc{R}),
\end{equation}
\end{widetext}
where $\Phi_{21}$ is the right hand side of \Eq{21expansionrealspace},
$l_{\text{max}}$ is the maximum possible orbital angular momentum quantum number of the two-body bound state,
and $(\nu_l+1)$ is the number of vibrational states for orbital angular momentum quantum number $l$.
\begin{equation}
    \varphi_{l\nu}(\vc{s},\vc{R})=u_{l\nu}(s)
    H_{l}^{(1)}\bigg(\frac{2}{\sqrt{3}}\kappa_{l\nu} R\bigg)
    \cos\left(l\,\theta\right),
\end{equation}
where $\theta$ is the angle between $\hat{\vc{s}}$ and $\hat{\vc{R}}$, and $u_{l\nu}(s)$ is the radial part of the dimer wave function for vibrational quantum number $\nu$ at
orbital angular momentum quantum number $l$, satisfying the two-body Schr\"odinger equation
\begin{align}
    &\left(-\nabla_\vect s^2+\kappa_{l\nu}^2\right)u_{l\nu}(s)\cos\left(l\,\theta\right)\nn\\
    &+\frac{1}{2}\int d^2s'U(\vc{s},\vc{s}')u_{l\nu}(s')\cos\left[l\,\text{ang}(\hat{\vc{s}}',\hat{\vc{R}})\right]=0.
\end{align}
$\text{ang}(\hat{\vc{s}}',\hat{\vc{R}})$ is the angle between $\hat{\vect s}'$ and $\hat{\vect R}$.
We impose the following normalization condition for $u_{l\nu}(s)$:
\begin{equation}
    \int_0^{\infty} s u_{l\nu}^*(s)u_{l\nu'}(s)ds=\frac{\delta_{\nu\nu'}}{\big(1+\delta_{l,0}\big)\pi\kappa_{l\nu}^2},
\end{equation}
where $\delta_{l,0}$ is the Kronecker delta.
From the conservation of probability, we derive a formula for the imaginary part of $D$:
\begin{equation}\label{ImD}
    \text{Im}{D}=-9\sum_{l\nu} \frac{|C_{l\nu}|^2}{\kappa_{l\nu}^2},
\end{equation}
which has the same form as Eq.~(118) in Ref. \cite{PhysRevA.109.063328}.
However, the $D$ in this paper has a very different meaning from the $D$ in Ref. \cite{PhysRevA.109.063328}:
the $D$ and $C_{l\nu}$ in this paper are defined for three bosons with a \emph{finite} two-body 2D scattering length $a$ and the $D$ in this paper appears at the order $1/(B^2\ln^3\widetilde{B})$ in the 111-expansion of the three-body wave function,
but the $D$ and $C_{l\nu}$ in Ref. \cite{PhysRevA.109.063328} are defined for three bosons with $a=\infty$ or $0$ and the $D$ in Ref. \cite{PhysRevA.109.063328} appears at the order $1/B^2$
in the 111-expansion of the three-body wave function.
If three bosons in two dimensions have a large but finite two-body 2D scattering length $a$, such that $a\gg r_e$,
then we anticipate that the three-body wave function has an expansion in the ``intermediate range" namely $r_e\ll B\ll a$
and another expansion in the ``large-distance limit" namely $B\gg a$,
the $D$ defined in Ref.~\cite{PhysRevA.109.063328} (denoted $D_{2024}$) should appear in the ``intermediate range" expansion,
the $D$ defined in this paper (denoted $D$) should appear in the ``large-distance limit" expansion,
and we conjecture that they have the following relation:
\beq
D\sim D_{2024}\ln^6(a/r_e).
\eeq
In other words, if we fine-tune the two-body interaction such that $a\to\infty$ or $0$, we expect that the $D$ defined in this paper
diverges like $\ln^6(a/r_e)$.

\section{Dependence of $D$ on the interaction potentials}
 
In this section, we investigate how small perturbations of the interaction potentials modify the two-body scattering parameters $\ln a$, $r_s$, and the three-body scattering parameter $D$. 

We assume that the two-body potential changes by an infinitesimal amount $\frac{\hbar^2}{m}\delta U(\vc{r}_1,\vc{r}_2, \vc{r}_1',\vc{r}_2')$ and that the three-body interaction potential changes by an infinitesimal amount $\frac{\hbar^2}{m}\delta U(\vc{r}_1,\vc{r}_2,\vc{r}_3,\vc{r}_1',\vc{r}_2',\vc{r}_3')$. Both $\delta U(\vc{r}_1,\vc{r}_2, \vc{r}_1',\vc{r}_2')$ and $\delta U(\vc{r}_1,\vc{r}_2,\vc{r}_3,\vc{r}_1',\vc{r}_2',\vc{r}_3')$ are assumed to satisfy the conditions specified in Sec.~\ref{sec:asymptotics}. For simplicity, we further assume that no two-body bound state is present.

The perturbations of the interaction potentials induce corresponding perturbations of the two-body special functions $\phi(\vc{s})$, $f(\vc{s})$, etc., and the zero-energy three-body wave function $\phi^{(3)}$. We write the resultant modified functions as $\phi'(\vc{s})$, $f'(\vc{s})$, etc., and $\phi'^{(3)}$. Consequently, the associated scattering parameters are modified: $\ln a\to \ln a +\delta \ln a$, $r_s\to r_s+\delta r_s$, $D\to D+\delta D$, etc. 

Using Eqs.~ (\ref{Hphi_l}), (\ref{Hf_l}), (\ref{phis,s>re}) and (\ref{f,s>re}), we find that
\begin{equation}\label{delta lna}
    \delta \ln a=\frac{\delta a}{a}=\frac{1}{4\pi}\int d^2s d^2s' \phi^*(s)\delta U(\vc{s},\vc{s}')\phi(\vc{s}')
\end{equation}
and
\begin{align}\label{delta rs}
    \delta r_s &= \frac{1}{\pi^2}\int d^2s d^2s'\Big[f^*(\vc{s})\delta U(\vc{s},\vc{s}')\phi(\vc{s}')\nonumber\\
    &\hspace{70pt}+\phi^*(\vc{s})\delta U(\vc{s},\vc{s}')f(\vc{s}')\Big].
\end{align}

To derive the expression for $\delta D$, we analyze the three-body Schr\"{o}dinger equations for $\phi^{(3)}$ and $\phi'^{(3)}$. 
We write the equations satisfied by $\phi^{(3)}$ and $\phi'^{(3)}$  as
\begin{equation}\label{H phi3=0}
    H^{(3)}\phi^{(3)}(\vc{r}_1,\vc{r}_2,\vc{r}_3)=0
\end{equation}
and 
\begin{align}\label{Hprime real space}
    &H'^{(3)}\phi'^{(3)}(\vc{r}_1,\vc{r}_2,\vc{r}_3)\nonumber\\
    \equiv &~ H^{(3)}\phi'^{(3)}(\vc{r}_1,\vc{r}_2,\vc{r}_3)\nonumber\\
    &+\frac{1}{2}\int d^2s_1' \delta U\left(\vc{s}_1,\vc{s}_1'\right)\phi'^{(3)}\left(\vc{R}_1,\frac{\vc{s}_1'}{2},-\frac{\vc{s}_1'}{2}\right)\nonumber\\
    &+\frac{1}{2}\int d^2s_2'\delta U\left(\vc{s}_2,\vc{s}_2'\right)\phi'^{(3)}\left(-\frac{\vc{s}_2'}{2},\vc{R}_2,\frac{\vc{s}_2'}{2}\right)\nonumber\\
    &+\frac{1}{2}\int d^2s_3'\delta U\left(\vc{s}_3,\vc{s}_3'\right)\phi'^{(3)}\left(\frac{\vc{s}_3'}{2},-\frac{\vc{s}_3'}{2},\vc{R}_3\right)\nonumber\\
    &+\frac{1}{6}\int dr_1'^2dr_2'^2 \delta U(\vc{r}_1,\vc{r}_2,\vc{r}_3,\vc{r}_1',\vc{r}_2',\vc{r}_3')\phi'^{(3)}(\vc{r}_1',\vc{r}_2',\vc{r}_3')\nonumber\\
    =&~0,
\end{align}
where $H^{(3)}$ and $H'^{(3)}$ denote the unperturbed and perturbed three-body Hamiltonians, respectively. The wave functions $\phi^{(3)}(\vc{r}_1,\vc{r}_2,\vc{r}_3)$ and $\phi'^{(3)}(\vc{r}_1,\vc{r}_2,\vc{r}_3)$ exhibit the asymptotic forms shown in \Eq{111expansion} and \Eq{21expansionrealspace}. 

From \Eq{H phi3=0} and \Eq{Hprime real space}, we get
\begin{align}\label{mixed H matrix element}
    \int d^2r_1d^2r_2\bigg[\phi'^{(3)*}(\vc{r}_1,\vc{r}_2,\vc{r}_3)H^{(3)}\phi^{(3)}(\vc{r}_1,\vc{r}_2,\vc{r}_3)\nonumber\\
    -\phi^{(3)}(\vc{r}_1,\vc{r}_2,\vc{r}_3)H'^{(3)*}\phi'^{(3)*}(\vc{r}_1,\vc{r}_2,\vc{r}_3)\bigg]=0.
\end{align}
Evaluating the left hand side of \Eq{mixed H matrix element}, using the divergence theorem, and using Eqs.~\eqref{111expansion}, \eqref{21expansionrealspace}, \eqref{delta lna}, and \eqref{delta rs}, we get
\begin{align}\label{delta D}
    \delta D&=\frac{3}{4}\pi^3\delta r_s\left[\left(\ln\frac{4}{3}\right)\mathrm{Li}_2\left(\frac{3}{4}\right)+2\mathrm{Li}_3\left(\frac{3}{4}\right)-3\zeta(3)\right]\nonumber\\
    &\quad +\frac{3}{2}\int d^2 s\int d^2s' \delta U(\vc{s},\vc{s}')\Xi(\vc{s},\vc{s}')\nonumber\\
    &\quad +\frac{1}{6}\int d^2r_1d^2r_2d^2r_1'd^2r_2' \phi^{(3)*}(\vc{r}_1,\vc{r}_2,\vc{r}_3)\nonumber\\
    &\quad\quad\quad\quad\delta U(\vc{r}_1,\vc{r}_2,\vc{r}_3,\vc{r}_1',\vc{r}_2',\vc{r}_3')\phi^{(3)}(\vc{r}_1',\vc{r}_2',\vc{r}_3'),
\end{align}
where $\vect r_3'\equiv\vect r_1+\vect r_2+\vect r_3-\vect r'_1-\vect r'_2$, and
\begin{align}
    &\quad\Xi(\vc{s},\vc{s}')\nonumber\\
    &= \lim_{\Lambda\to \infty}\int_{R<\Lambda} d^2 R \,\phi^{(3)*}\left(\vc{R},\frac{\vc{s}}{2},-\frac{\vc{s}}{2}\right)\phi^{(3)}\left(\vc{R},\frac{\vc{s}'}{2},-\frac{\vc{s}'}{2}\right)\nonumber\\
    &\quad-\phi(\vc{s})^*\phi(\vc{s}')\pi \Lambda^2\Big[\ln^4\widetilde{\Lambda}-2\ln^3\widetilde{\Lambda}\nonumber\\
    &\quad\quad\quad+(3-4\omega)\left(\ln^2\widetilde{\Lambda}-\ln\widetilde{\Lambda}\right)+4\omega^2-2\omega+\frac{3}{2}\Big]\nonumber\\
    &\quad-\left[\phi^*(\vc{s})f(\vc{s}')+\phi(\vc{s}')f^*(\vc{s})\right]\pi(\ln^3\widetilde{\Lambda}-6\omega \ln\widetilde{\Lambda}),
\end{align}
where $\widetilde{\Lambda}=\frac{8\Lambda}{3\sqrt{3}a}$.

If we slightly change the three-body potential, but keep the two-body potential unchanged, Eq.~(\ref{delta D}) is simplified as
\begin{align}\label{delta D,2}
    \delta D&=\frac{1}{6}\int d^2r_1d^2r_2d^2r_1'd^2r_2' \phi^{(3)*}(\vc{r}_1,\vc{r}_2,\vc{r}_3)\nonumber\\
    &\quad\quad\quad\quad\delta U(\vc{r}_1,\vc{r}_2,\vc{r}_3,\vc{r}_1',\vc{r}_2',\vc{r}_3')\phi^{(3)}(\vc{r}_1',\vc{r}_2',\vc{r}_3').
\end{align}

\section{Implications for the three-body and the many-body physics}

Let us consider any number of identical bosons in a large square with side length $L\gg\max(r_e,a)$, and impose the periodic boundary condition such that $\psi(\vect r+L\vect n)=\psi(\vect r)$, where $\psi(\vc{r})$ is the boson annihilation operator at position $\vect r$, and $\vect n=n_x\hat{\vect x}+n_y\hat{\vect y}$ is any vector with integer Cartesian components $n_x$ and $n_y$.
$\psi(\vect r)$ satisfies $\left[\psi(\vc{r}),\psi(\vc{r}')\right]=0$ 
and $\left[\psi(\vc{r}),\psi^\dagger(\vc{r}')\right]=\sum_\vect n\delta(\vc{r}-\vc{r}'-L\vect n)$. Let $|\text{vac}\rangle$ be the normalized vacuum state satisfying $\psi(\vc{r})|\text{vac}\rangle=0$. 
The second-quantized Hamiltonian with two-body potential $\frac{\hbar^2}{m}U(\vc{r}_1,\vc{r}_2,\vc{r}_1',\vc{r}_2')$ and three-body potential $\frac{\hbar^2}{m}U(\vc{r}_1,\vc{r}_2,\vc{r}_3,\vc{r}_1',\vc{r}_2',\vc{r}_3')$, which satisfy the assumptions in Sec. \ref{sec:asymptotics}, is
 \begin{widetext}
 \begin{align}
     \widehat{H}=&~\frac{\hbar^2}{2m}\int d^2r\nabla\psi^\dagger(\vc{r})\cdot\nabla\psi(\vc{r})\nonumber\\
     &~+\frac{\hbar^2}{m} \int d^2r_1 d^2r_2 d^2r_1' d^2r_2' U(\vc{r}_1,\vc{r}_2,\vc{r}_1',\vc{r}_2')\delta(\vc{r}_1+\vc{r}_2-\vc{r}_1'-\vc{r}_2')\psi^{\dagger}\left(\vc{r}_1\right)\psi^{\dagger}\left(\vc{r}_2\right)\psi\left(\vc{r}_2'\right)\psi\left(\vc{r}_1'\right)\nonumber\\
     &~+\frac{\hbar^2}{36m} \int d^2r_1 d^2r_2 d^2r_3 d^2r_1' d^2r_2' d^2r_3' U\left(\vc{r}_1,\vc{r}_2,\vc{r}_3,\vc{r}_1',\vc{r}_2',\vc{r}_3'\right)\delta(\vc{r}_1+\vc{r}_2+\vc{r}_3-\vc{r}_1'-\vc{r}_2'-\vc{r}_3')\nonumber\\
    &~\hspace{21em}\psi^{\dagger}\left(\vc{r}_1\right)\psi^{\dagger}\left(\vc{r}_2\right)\psi^{\dagger}\left(\vc{r}_3\right)\psi\left(\vc{r}_3'\right)\psi\left(\vc{r}_2'\right)\psi\left(\vc{r}_1'\right),
 \end{align}
 \end{widetext}
 where the integral over each vector is carried out in a single square with side length $L$.
We consider an $N$-body energy eigenstate
\begin{align}\label{ket N body}
    |\psi_N\rangle = \frac{1}{\sqrt{N!}}\int& d^2r_1 \cdots d^2r_N \psi_N(\vc{r}_1,\cdots,\vc{r}_N)\nonumber\\
    &\times\psi^\dagger(\vc{r}_1)\cdots\psi^\dagger(\vc{r}_N)|\text{vac}\rangle
\end{align}
with energy $E$, where $\psi_N(\vc{r}_1,\ldots,\vc{r}_N)$ is the $N$-body wave function.
$\psi_N$ is symmetric under the exchange of $\vect r_i$
and $\vect r_j$ for $1\le i<j\le N$, satisfies the periodic boundary condition, and is normalized such that $\int  |\psi_N|^2\prod_{i=1}^N d^2r_i=1$. Thus, $\langle \psi_N | \psi_N \rangle=1$.

As we \emph{adiabatically} change the three-body potential from $U(\vect r_1,\vect r_2,\vect r_3,\vect r_1',\vect r_2',\vect r_3')$ to $U(\vect r_1,\vect r_2,\vect r_3,\vect r_1',\vect r_2',\vect r_3')+\delta U(\vect r_1,\vect r_2,\vect r_3,\vect r_1',\vect r_2',\vect r_3')$,
where $\delta U(\vect r_1,\vect r_2,\vect r_3,\vect r_1',\vect r_2',\vect r_3')$ is infinitesimal, but keep the two-body potential unchanged,
the first-order energy shift  of the above $N$-body energy level is
 \begin{align}\label{delta E general}
     \delta E=~&\frac{\hbar^2}{36m}\int\delta U(\vc{r}_1,\vc{r}_2,\vc{r}_3,\vc{r}_1',\vc{r}_2',\vc{r}_3')\nonumber\\
     ~&\hspace{-2em}\delta(\vc{r}_1+\vc{r}_2+\vc{r}_3-\vc{r}_1'-\vc{r}_2'-\vc{r}_3')g_3\left(\vc{r}_1,\vc{r}_2,\vc{r}_3,\vc{r}_3',\vc{r}_2',\vc{r}_1'\right)\nonumber\\
     ~& d^2r_1d^2r_2d^2r_3d^2r_1'd^2r_2'd^2r_3',
 \end{align}
 where
 \begin{align}\label{g3}
     &g_3\left(\vc{r}_1,\vc{r}_2,\vc{r}_3,\vc{r}_3',\vc{r}_2',\vc{r}_1'\right)\nonumber\\
     &=\left\langle\psi_N\left|\psi^{\dagger}\left(\vc{r}_1\right)\psi^{\dagger}\left(\vc{r}_2\right)\psi^{\dagger}\left(\vc{r}_3\right)\psi\left(\vc{r}_3'\right)\psi\left(\vc{r}_2'\right)\psi\left(\vc{r}_1'\right)\right|\psi_N\right\rangle\nn\\
     &=N(N-1)(N-2)\int d^2r_{4}\cdots d^2r_N \nonumber\\
    &\quad\times\psi_N^*\left(\vc{r}_1,\vc{r}_2,\vc{r}_3,\vc{r}_{4},\ldots,\vc{r}_N\right)\psi_N\left(\vc{r}_1',\vc{r}_2',\vc{r}_3',\vc{r}_{4},\ldots,\vc{r}_N\right)
 \end{align}
is the nonlocal three-body correlation function.

\subsection{Shift of the three-body ground state energy}
Let us first consider the ground state of three identical bosons (assuming that there is no two-body or three-body bound state) with normalized wave function $\psi_3(\vc{r}_1,\vc{r}_2,\vc{r}_3)$. For $L\gg\max(r_e,a)$, $\psi_3$ is
approximately a constant when the pairwise distances are all comparable to $L$, and  the normalized wave function may be written as
\begin{equation}\label{three boson wf in square,si L}
    \psi_3(\vc{r}_1,\vc{r}_2,\vc{r}_3)\approx L^{-3},\quad\text{if } s_1,s_2,s_3\sim L.
\end{equation}
On the other hand, when $B\ll L$, $\psi_3$ can be approximated by some coefficient $\mu$ times the dominant three-body zero-energy wave function with zero orbital angular momentum:
\begin{equation}\label{three boson wf in square, B<<L}
    \psi_3(\vc{r}_1,\vc{r}_2,\vc{r}_3)\approx \mu
    \phi^{(3)}(\vc{r}_1,\vc{r}_2,\vc{r}_3),~  B\ll L,
\end{equation}
where $\phi^{(3)}(\vc{r}_1,\vc{r}_2,\vc{r}_3)$ has the asymptotic form shown in \Eq{111expansion} at large pairwise distances.
Since the two approximate forms of $\psi_3(\vc{r}_1,\vc{r}_2,\vc{r}_3)$ shown in \Eq{three boson wf in square,si L} and \Eq{three boson wf in square, B<<L} should be consistent, and since when $s_1, s_2, s_3 \gg \max(r_e,a)$, $\phi^{(3)}(\vc{r}_1,\vc{r}_2,\vc{r}_3)\approx \ln^3\frac{B}{a}$ (assuming that $\ln\frac{B}{a}\gg1$), we have
\begin{equation}
 \mu\approx\frac{1}{L^3\ln^3\frac{L}{a}}.
\end{equation}
The three-body correlation function is thus
\begin{align}\label{g3 three body}
    &~g_3\left(\vc{r}_1,\vc{r}_2,\vc{r}_3,\vc{r}_3',\vc{r}_2',\vc{r}_1'\right)\nonumber\\
    &=\nonumber6\psi_3^{*}(\vc{r}_1,\vc{r}_2,\vc{r}_3)\psi_3(\vc{r}_1',\vc{r}_2',\vc{r}_3')\\
    &\approx6\phi^{(3)*}(\vc{r}_1,\vc{r}_2,\vc{r}_3)\phi^{(3)}(\vc{r}_1',\vc{r}_2',\vc{r}_3')\Big/\left(L^6\ln^6\frac{L}{a}\right)
\end{align}
at $B,B'\ll L$, where 
$$B'\equiv\sqrt{\frac12(|\vect r_2'-\vect r_3'|^2+|\vect r_3'-\vect r_1'|^2+|\vect r_1'-\vect r_2'|^2)}.$$
Substituting \Eq{g3 three body} into \Eq{delta E general} and using \Eq{delta D,2}, we get
\begin{align}\label{delta E 3-body}
    \delta E\approx\frac{\hbar^2\delta D}{mL^4\ln^{6}\frac{L}{a}}.
\end{align}
We believe that if one slightly changes the two-body potential in a particular way such that 
the two-body scattering length $a$ and effective range $r_s$ do \emph{not} change [namely, if one chooses $\delta U(\vect s,\vect s')$ such that the $\delta\ln a$ and $\delta r_s$ shown in Eqs.~\eqref{delta lna} and \eqref{delta rs} both vanish],
\Eq{delta E 3-body} is still valid.

\subsection{Shift of the many-body energy in the thermodynamic limit and its implications}
Next, we consider the case of $N$ bosons in the thermodynamic limit, such that
the number density $\rho\equiv N/L^2$ is fixed, but $N\to\infty$ and $L\to \infty$.
Further, we consider the dilute limit,  such that the interparticle spacing $d=\rho^{-\frac{1}{2}}\gg \max(a,r_e)$. In this limit, the ground state energy in the absence of bound states can be expanded in terms of the small parameter $\rho a^2$ \cite{schick1971two,popov1972theory,lozovik1978ground,fisher1988dilute,kolomeisky1992renormalization,ovchinnikov1993description,lieb2001rigorous,al2002low,pricoupenko2004variational,pilati2005quantum,mora2009ground,pastukhov2019ground,fournais2024ground,PhysRevA.81.013612,beane2010ground,PhysRevLett.118.130402,beane2018effective,Tononi_2018,condmat4010020,PhysRevA.107.033325}. 
To analyze finite-temperature effects,
we define the quantum degeneracy temperature
\beq
T_d \equiv \frac{2\pi\hbar^2\rho}{m k_B},
\eeq
where $k_B$ is the Boltzmann constant. We also define the few-body temperature scales
\beq
T_e \equiv \frac{2\pi\hbar^2}{m r_e^2 k_B},
\eeq
\beq
T_a \equiv \frac{2\pi\hbar^2}{m a^2 k_B}.
\eeq

If the temperature $T$ of the Bose gas satisfies $T\ll\min(T_e,T_a)$,
the thermal de Broglie wave length 
\beq
\lambda_T \equiv \sqrt{\frac{2\pi\hbar^2}{m k_BT}}
\eeq
is much longer than $\max(r_e,a)$, and we may assume that if three of the $N$ bosons are close to each other but the remaining $(N-3)$ bosons are not close to them, the wave function of a typical $N$-body energy eigenstate is approximately proportional to the function $\phi^{(3)}$ studied in Sec.~\ref{sec:asymptotics}, with a constant of proportionality that depends on the center-of-mass position of the three bosons as well as the positions of the remaining $(N-3)$ bosons:
\begin{align}\label{N body wf factor}
    &~\psi_N(\vc{r}_1,\vc{r}_2,\vc{r}_3,\vc{r}_4,\cdots,\vc{r}_N)\nonumber\\
    &\approx G(\vc{c}_{123},\vc{r}_4,\cdots,\vc{r}_N)\phi^{(3)}(\vc{r}_1,\vc{r}_2,\vc{r}_3),
\end{align}
if $B\ll \min(d,\lambda_T)$ and $|\vc{r}_i-\vc{c}_{123}|\gg\max(B,a,r_e)$ for $4\le i\le N$.
Here $B\equiv\sqrt{\frac12(s_1^2+s_2^2+s_3^2)}$ is the hyperradius of the triangle formed by $\vect r_1$, $\vect r_2$, and $\vect r_3$,
and $\vc{c}_{123}\equiv \left(\vc{r}_1+\vc{r}_2+\vc{r}_3\right)/3$ is the center-of-mass position vector of bosons 1, 2, and 3.

In \Eq{g3}, if the hyperradii $B\equiv\sqrt{\frac12(s_1^2+s_2^2+s_3)}$
(where $\vect s_1\equiv\vect r_2-\vect r_3$, $\vect s_2\equiv\vect r_3-\vect r_1$, and $\vect s_3\equiv\vect r_1-\vect r_2$) 
and $B'\equiv \sqrt{\frac12(s_1'^2+s_2'^2+s_3'^2)}$
(where $\vect s_1'\equiv\vect r_2'-\vect r_3'$, $\vect s_2'\equiv\vect r_3'-\vect r_1'$, and $\vect s_3'\equiv\vect r_1'-\vect r_2'$) are both much smaller than $\min(d,\lambda_T)$,
the contributions to the $2(N-3)$-fold integral from those cases in which
$\vect r_4$, $\vect r_5$, \dots, or $\vect r_N$ is close to $\vect c_{123}$
or $\vect c'_{123}$
[where $\vc{c}_{123}'\equiv \left(\vc{r}_1'+\vc{r}_2'+\vc{r}_3'\right)/3$]
can be neglected.
Therefore, if we assume that the $N$ bosons are in a total momentum eigenstate, we can
substitute \Eq{N body wf factor} into \Eq{g3} to find
\begin{align}\label{g3 factor}
    &g_3\left(\vc{r}_1,\vc{r}_2,\vc{r}_3,\vc{r}_3',\vc{r}_2',\vc{r}_1'\right)\nonumber\\
    \approx~&N(N-1)(N-2)\phi^{(3)*}(\vc{r}_1,\vc{r}_2,\vc{r}_3)\phi^{(3)}(\vc{r}_1',\vc{r}_2',\vc{r}_3')\nonumber\\
    ~&\hspace{-1em}\int d^2r_{4}\cdots d^2r_N G^*(\vc{c}_{123},\vc{r}_4,\cdots,\vc{r}_N)G\left(\vc{c}_{123}',\vc{r}_4,\cdots,\vc{r}_N\right)\nonumber\\
    =~&\eta(\vect c_{123}-\vect c_{123}')\phi^{(3)*}(\vc{r}_1,\vc{r}_2,\vc{r}_3)\phi^{(3)}(\vc{r}_1',\vc{r}_2',\vc{r}_3')
\end{align}
at $B,B'\ll\min(d,\lambda_T)$,
where the coefficient $\eta$ is a function of $\vect c_{123}-\vect c_{123}'$. 
The rotational symmetry of the interactions implies that if we take the thermal-ensemble
average of $g_3$ for all energy eigenstates within the ensemble with temperature $T$, we should get a function $\eta$ that does \emph{not} depend on the direction of $\vect c_{123}-\vect c_{123}'$.

Substituting \Eq{g3 factor} into \Eq{delta E general}, and using \Eq{delta D,2}, we derive
the adiabatic shift of the many-body energy density:
 \begin{align}\label{delta E thermodynamic}
     \frac{\delta E}{L^2}\approx \frac{\hbar^2 \eta(\vc{0})\delta D}{6m},
 \end{align}
 which should be valid if $T\ll\min(T_e,T_a)$.
We believe that if one slightly changes the two-body potential in a particular way such that 
the two-body scattering length $a$ and effective range $r_s$ do \emph{not} change [namely, if one chooses $\delta U(\vect s,\vect s')$ such that the $\delta\ln a$ and $\delta r_s$ shown in Eqs.~\eqref{delta lna} and \eqref{delta rs} both vanish],
\Eq{delta E thermodynamic} is still valid.

 If the two-body interaction supports two-body bound states,  $D$ usually has a negative imaginary part shown in \Eq{ImD}.
We expect that \Eq{delta E thermodynamic} remains applicable in this case, so that now the energy $E$ has a negative imaginary part proportional to $|\mathrm{Im}D|$, implying that the many-body state is now metastable.
Assuming that the two-body parameters such as $a$ and $r_s$ are real, from \Eq{delta E thermodynamic} we get $|\mathrm{Im}E|\approx \frac{\hbar^2 \eta(\vc{0})L^2|\mathrm{Im} D| }{6m}$.
Within a short time interval $dt$, the probability that no three-body recombination occurs is $\exp(-2 |\mathrm{Im}E|dt/\hbar)\approx 1-2 |\mathrm{Im}E|dt/\hbar$, and so the probability of one recombination is $2 |\mathrm{Im}E|dt/\hbar$.  Each recombination results in the loss of three low-energy bosons, so
\begin{equation}
    \frac{d\rho}{dt}=-L_3 \rho^3,
\end{equation}
where the three-body recombination rate constant
\beq\label{L3}
L_3\approx\frac{\hbar\eta(\vc{0}) |\mathrm{Im} D|}{m\rho^3}.
\eeq

Next, we discuss the temperature dependence of the coefficient $\eta(\vect 0)$
and the implications for $L_3$.

At $T=0$, the Bose gas forms a Bose-Einstein condensate. If $\rho a^2$ is so small that $|\ln(\rho a^2)|\gg1$, almost $100\%$ of the bosons are in the zero-momentum condensed state \cite{schick1971two}. 
At temperatures $0<T\ll T_{\mathrm{BKT}}$, where $T_{\mathrm{BKT}}$ is the Berezinskii--Kosterlitz--Thouless transition temperature, 
$T_{\mathrm{BKT}} \approx \frac{2\pi\hbar\rho}{m k_B}
\frac{1}{\ln 30.2 + \ln\!\ln(1/\rho a^2)}$ \cite{PhysRevLett.87.270402,PhysRevLett.100.140405}, a true condensate is absent, but a quasicondensate persists
\cite{popov1972theory}. 
Therefore, if the pairwise distances $s_1$, $s_2$, $s_3$, $s_1'$, $s_2'$, and $s_3'$ are all comparable to $d$ or somewhat smaller than $d$, and $\vect c_{123}=\vect c'_{123}$,
the three-body correlation function
\begin{equation}\label{g3 appr 1}
g_3\!\left(\vc{r}_1,\vc{r}_2,\vc{r}_3,
\vc{r}_3',\vc{r}_2',\vc{r}_1'\right)
\approx \rho^3
\end{equation}
as shown in Refs.~\cite{1987JETP66314K,PhysRevA.61.043608,PhysRevA.66.043608}.
On the other hand, according to Eq.~\eqref{g3 factor} and \Eq{111expansion}, if $s_1$, $s_2$, $s_3$, $s_1'$, $s_2'$, and $s_3'$
are all somewhat smaller than $d$, and $\vect c_{123}=\vect c'_{123}$,
\begin{align}\label{g3 appr 2}
g_3\left(\vc{r}_1,\vc{r}_2,\vc{r}_3,\vc{r}_3',\vc{r}_2',\vc{r}_1'\right)
&\approx \eta(\vect 0) \ln^6\frac{d}{a},\nonumber\\
&=\frac{1}{64}\eta(\vect 0) \ln^6(\rho a^2).
\end{align} 
Comparing Eq.~\eqref{g3 appr 1} with Eq.~\eqref{g3 appr 2}, we find
\begin{equation}\label{eta result}
\eta(\vect 0)\approx \frac{64\rho^3}{\ln^6(\rho a^2)}
\end{equation}
if $T\ll T_{\mathrm{BKT}}$. Substituting \Eq{eta result} into \Eq{L3}, we get
\beq\label{L3,T<<Tc}
L_3\approx\frac{64\hbar|\mathrm{Im} D|}{m\ln^6\left(\rho a^2\right)}
\eeq
at $T\ll T_\mathrm{BKT}$.
Ref.~\cite{1987JETP66314K} also found the relation $L_3 \propto \ln^{-6}(\rho a^2)$
, while we have got the above closed-form formula for $L_3$ in terms of the imaginary part of the three-body scattering area $D$.

If $T\gg T_\mathrm{BKT}$ but $T$ is less than, similar to, or \emph{not} much greater than $T_d$, the bosonic bunching becomes important.
For $s_1$, $s_2$, $s_3$, $s_1'$, $s_2'$, and $s_3'$ all somewhat smaller than $d$, the three-body correlation function is enhanced to
$g_3\!\left(\vc{r}_1,\vc{r}_2,\vc{r}_3,
\vc{r}_3',\vc{r}_2',\vc{r}_1'\right)
\approx 6\rho^3$
\cite{1987JETP66314K,PhysRevA.61.043608}. Meanwhile, when $s_1$, $s_2$, $s_3$, $s_1'$, $s_2'$, and $s_3'$
are all somewhat smaller than $d$, and $\vect c_{123}=\vect c'_{123}$, $g_3$ retains the form shown in \Eq{g3 appr 2}, so we get
\begin{equation}
    \eta(\vc{0})\approx \frac{384\rho^3}{\ln^6(\rho a^2)},
\end{equation}
\beq
L_3\approx\frac{384\hbar|\mathrm{Im} D|}{m\ln^6\left(\rho a^2\right)}.
\eeq

We next consider the even higher temperature regime
$T_d \ll T \ll \min\left(T_e, T_a\right)$, in which $\max(r_e,a)\ll\lambda_T\ll d$.
If $\vect c_{123}=\vect c'_{123}$ and the pairwise distances satisfy
$\max\left(r_e, a\right) \ll s_1, s_2, s_3, s_1', s_2', s_3'
\ll \lambda_T$,
bosonic bunching again enhances the three-body correlation function to
$g_3\!\left(\vc{r}_1,\vc{r}_2,\vc{r}_3,
\vc{r}_3',\vc{r}_2',\vc{r}_1'\right)
\approx 6\rho^3$
\cite{1987JETP66314K,PhysRevA.61.043608}. On the other hand, according to Eq.~\eqref{g3 factor} and \Eq{111expansion}, when $s_1$, $s_2$, $s_3$, $s_1'$, $s_2'$, and $s_3'$
are all somewhat smaller than $\lambda_T$, and $\vect c_{123}=\vect c'_{123}$,
\begin{equation}\label{g3 appr 3}
g_3\left(\vc{r}_1,\vc{r}_2,\vc{r}_3,\vc{r}_3',\vc{r}_2',\vc{r}_1'\right)
\approx \eta(\vect 0) \ln^6(\lambda_T/a).
\end{equation} 
So we get
\begin{equation}\label{eta, Td<<T<<Ta}
    \eta(\vc{0})\approx \frac{6\rho^3}{\ln^6(\lambda_T/a)},
\end{equation}
\beq\label{L3,T<<Tc, 2nd}
L_3\approx\frac{6\hbar|\mathrm{Im} D|}{m\ln^6\left(\lambda_T/a\right)}.
\eeq
The relation $L_3\propto \ln^{-6}\frac{\lambda_T}{a}$  can also be obtained from the results of Ref.~\cite{PhysRevA.91.062710}.

\begin{acknowledgments}
This work was supported by the National Key R$\&$D Program of China Grant No. 2021YFA1400902,
the National Natural Science Foundation of China Grant No. 92365202,
and the National Key R$\&$D Program of China Grant No. 2019YFA0308403.
We thank Zipeng Wang and Jiansen Zhang for helpful discussions.
\end{acknowledgments}

\bibliography{apssamp}
\end{document}